\documentclass[11pt,a4paper]{article} 
\pdfoutput=1

\usepackage{style}

\usepackage{amsmath,amssymb,epsfig,bm,amsfonts,yfonts}

\usepackage{subfigure}
\relax
\title{Causality of fluid dynamics for high-energy nuclear collisions}

\author{Stefan Floerchinger}
\author{and Eduardo Grossi}
\affiliation{Institut f\"{u}r Theoretische Physik, Universit\"{a}t Heidelberg, Philosophenweg 16, 69120 Heidelberg, Germany}
\emailAdd{stefan.floerchinger@thphys.uni-heidelberg.de}
\emailAdd{e.grossi@thphys.uni-heidelberg.de}

\abstract{
Dissipative relativistic fluid dynamics is not always causal and can favor superluminal signal propagation under certain circumstances. On the other hand, high-energy nuclear collisions have a microscopic description in terms of QCD and are expected to follow the causality principle of special relativity. 
We discuss under which conditions the fluid evolutions for a radial expansion are hyperbolic and how the properties of the solutions are encoded in the associated characteristic curves. 
The expansion dynamics is causal in relativistic sense if the characteristic velocities are smaller than the speed of light. 
We obtain a concrete inequality from this constraint and discuss how it can be violated for certain initial conditions. We argue that causality poses a bound to the applicability of relativistic fluid dynamics. }
\begin{document}
\maketitle
\section{Introduction}
\label{int}
Relativistic fluid dynamics has been used with success as an effective theory of QCD in the regime of high energy density probed by relativistic heavy ion collisions \cite{Heinz:2013th, Gale:2013da, Huovinen:2013wma, deSouza:2015ena, Schafer:2009dj}. It is seen as an expansion around local thermal equilibrium states in terms of derivatives. The lowest order in this derivative expansion is ideal fluid dynamics and describes in particular equilibrium states. Higher orders in a formal derivative expansion describe viscous and other transport corrections. The microscopic physics of QCD enters the formalism in terms of the thermodynamic equation of state and the transport properties such as viscosities, conductivities or relaxation times.

In  recent years the classical derivation of relativistic fluid dynamics from kinetic theory \cite{Muller:1967zza, Israel:1979wp, Muronga:2003ta, Denicol:2012cn} -- which  is valid for weakly coupled (perturbative) quantum field theories  -- has been supplemented  with a derivation \cite{Baier:2007ix, Bhattacharyya:2008jc, Policastro:2001yc, Kovtun:2004de, Romatschke:2009kr, Haehl:2014zda, Haehl:2015pja} from the non-perturbative setup of the AdS/CFT correspondence \cite{Maldacena:1997re}. Further works have addressed the convergence properties of the Chapman-Enskog gradient expansion around local equilibrium  (formulated in terms of equilibrium fields such as temperature and fluid velocity) and argued that relativistic fluid dynamics  could be understood as a resummation of this Chapman-Enskog expansion \cite{Heller:2015dha,Denicol:2016bjh, Florkowski:2017olj,Romatschke:2017acs, Romatschke:2017ejr}.

One of the most puzzling properties of relativistic fluid dynamics is that it is not always causal. More specific, it has been recognized early that relativistic versions of the Navier-Stokes equation, formulated either in the Landau frame or in the Eckart frame, allow for signal propagation with arbiltrarilly  large velocity. This can be traced back to the fact that the non-relativistic Navier-Stokes equation is a parabolic equation (and not hyperbolic as for example the Klein-Gordon equation). Hyperbolic extensions of this setup have been proposed, most prominently  by Israel and Stewart \cite{Israel:1979wp} (see also refs.\ \cite{Kostadt:2000ty, Herrera:2001if, Kostadt:2001rr, Muronga:2003ta} for related discussions). The setup of Israel and Stewart goes beyond the Chapman-Enskog expansion and shear stress, bulk viscous pressure and diffusion current are dynamic fields with their own evolution equations.
It was subsequently shown by Hiscock and Lindblom \cite{Hiscock:1983zz} that for appropriate values of the velocity of sound and transport properties, the theory of Israel and Stewart has equilibrium states around which small perturbations evolve in a relativistically causal way, indeed. They also investigated the stability of these linear perturbations and found that the conditions for causality also lead to linear stability of the homogeneous  equilibrium states.

A generalization of Isreal-Stewart theory that is more complete in the sense that it takes all possible terms at second order in derivatives into account, has been put forward by Denicol, Niemi, Molnar and Rischke \cite{Denicol:2012cn}; we will review their results in section \ref{sec2}. The relation between causality and linear stability for perturbations around equilibrium states was investigated in more detail in refs.\ \cite{Denicol:2008ha, Pu:2009fj}. It was confirmed that causality, in the sense of an asymptotic group velocity that is smaller than the speed of light, implies linear stability.

So far, investigations of causality were restricted to small perturbations of thermal equilibrium states. It is certainly necessary that such perturbations behave in a causal way, however, it is not sufficient to guarantee causality in more general situations. In fact, in the absence of a more general theoretical argument, causality needs to be established for every solution of the non-linear relativistic fluid equations of motion, case by case. This might be different only if the theory is organized in a scheme different from the conventional derivative expansion, for example as in ref.\ \cite{Geroch:1990bw}. 

In the present work, we will consider a class of such solutions of relativistic fluid dynamics describing a fireball produced by high-energy nuclear collisions with longitudinal and transverse expansion. We discuss the corresponding solution of the equations proposed in ref.\ \cite{Denicol:2012cn} and we discuss the issue of causality. Our main result will be concrete relations in the form of inequalities that tell whether perturbations propagating in the radial direction around the full, non-linear solutions of the fluid dynamic equations, behave in a causal way. This provides a non-linear form of the causality constraint and goes beyond the causality constraint for perturbations around equilibrium states. In particular, it turns out that the causality bounds are not satisfied for arbitrary initial conditions. We discuss what kind of initial conditions are save in this respect.

Throughout the manuscript we will work with natural units where $c=\hbar=k_B=1$ and we will take the signature of the Minkowski metric to be $(-,+,+,+)$.
\section{Hyperbolic partial differential equations and causality}
\label{eq:HyperbolicPDEs}

\subsection{Sets of hyperbolic partial differential equations}
In the present section we recall some mathematical knowledge about partial differential equations of the type relevant for dissipative relativistic fluid dynamics. This will lay the ground for a proper discussion of causality. We follow here mainly ref.\ \cite{1953mmp..book.....C}.

Let us consider a set of first order partial differential equations of the type
\begin{equation}
\label{eq:hyperbolic}
A_{ij}(\Phi)\;\frac{\partial}{\partial x^0} \Phi_j+ B_{ij}(\Phi) \;\frac{\partial}{\partial x^1} \Phi_j +C_i(\Phi)=0.
\end{equation}
We will see that the fluid equations of motion for a longitudinally and radially expanding fluid can be expressed in this form under rather general conditions. The independent variables or coordinates are a time variable $x^0$ and a spatial (radial) variable $x^1$. The dependent variables or fields are collected into the vector structure $\Phi_j$ with index $j=1, \ldots, n$. The coefficient matrices $A_{ij}$, $B_{ij}$ as well as the ``source terms'' $C_i$ are functions of the fields $\Phi$ and of the coordinates $(x^0,x^1)$ but do not depend on derivatives of $\Phi$.

Let us recall that the system of partial differential equations \eqref{eq:hyperbolic} would be called linear if $A_{ij}$ and $B_{ij}$ were independent of $\Phi$ and $C_i$ would depend at most linearly on $\Phi$. If $A_{ij}$ and $B_{ij}$ were independent of $\Phi$ and the source terms $C_i$ some (non-linear) function of $\Phi$, the system would be called semi-linear. In the more general case (relevant to us) where $A_{ij}$ and $B_{ij}$ are non-linear functions of $\Phi$ one calls \eqref{eq:hyperbolic} a quasi-linear set of partial differential equations of first order. The equations \eqref{eq:hyperbolic} would be called  homogeneous for $C_i(\Phi)=0$.

We also assume that the matrix $A_{ij}$ in front of the time derivative is non-singular, $\det(A)\neq 0$, and can be inverted. In the following we consider Cauchy's initial value problem, formulated as follows. We start with some initial configuration on some curve ${\cal C}$. This curve represents a Cauchy surface in the case of 1+1 dimensions and will in practice be for example a line of constant time $x^0=\text{const}$. More general, we assume that the curve $\cal C$ is specified by an equation $\varphi(x^0,x^1)=0$ with $\partial_0 \varphi(x^0, x^1) \neq 0$. We now ask whether the information given in terms of the field values $\Phi_j(x)$ on the curve $\cal C$ is sufficient to determine the first derivatives of $\Phi_j$ via the system of equations \eqref{eq:hyperbolic}.

We first note that with $\Phi_j(x)$ on $\cal C$, we have also all the information to determine the internal derivatives\footnote{A derivative operator $\alpha^0 \partial_0 + \alpha^1 \partial_1$ is called internal with respect to the curve specified by $\varphi(x)=0$ if $[\alpha^0 \partial_0 + \alpha^1 \partial_1] \varphi =0$. This is obviously the case for $\alpha^0=-\partial_1 \varphi$ and $\alpha^1=\partial_0 \varphi$.} or derivatives along $\cal C$
\begin{equation}
-(\partial_1 \varphi) \partial_0 \Phi_j + (\partial_0 \varphi) \partial_1 \Phi_j = (\partial_0 \varphi) D_j,
\end{equation}
i.\ e.\ we can take $D_j$ to be also known along $\cal C$. We can solve this equation for $\partial_1\Phi_j$ and, using $\lambda = - \partial_1 \varphi / \partial_0 \varphi$, write \eqref{eq:hyperbolic} as
\begin{equation}
\left[ A_{ij}(\Phi) - \lambda \, B_{ij}(\Phi) \right] \partial_0 \Phi_j + C_i(\Phi) + B_{ij}(\Phi) D_j =0.
\end{equation}
This is now a linear set of equations for the time derivatives $\partial_0 \Phi_j$. Accordingly, a necessary and sufficient condition for the first derivatives to be uniquely determined by \eqref{eq:hyperbolic} along the curve $\cal C$ is given by
\begin{equation}
Q = \det \left[ A_{ij}(\Phi) - \lambda \, B_{ij}(\Phi) \right] \neq 0.
\end{equation}
Here, $Q$ is known as the characteristic determinant. If $Q\neq 0$ along the curve $\cal C$, the latter is called free. The fields $\Phi_j$ along such curves can be continued into a ``strip'' where they solve \eqref{eq:hyperbolic}. By going in this way from one Cauchy surface to the next, one can construct solutions of \eqref{eq:hyperbolic}.

If $\lambda^{(m)}$ is a real solution of the algebraic equation $Q=0$, the solution of the differential equation
\begin{equation}
\frac{d x^1}{d x^0} = \lambda^{(m)},
\end{equation}
is called a {\it characteristic curve}. The real eigenvalue $\lambda^{(m)}$ is known as a {\it characteristic velocity}. Note that characteristic curves are {\it not} free in the sense defined above.  For a given real solution $\lambda^{(m)}$ of $Q=0$ one can find a corresponding (left) eigenvector $w^{(m)}_i$ such that
\begin{equation}
w_i^{(m)} \left[ A_{ij}(\Phi) - \lambda^{(m)} \, B_{ij}(\Phi) \right] = 0.
\end{equation}
The system of $n$ equations \eqref{eq:hyperbolic} is called hyperbolic if $n$ linearly independent eigenvectors $w^{(m)}_i$ with corresponding real eigenvectors $\lambda^{(m)}$ can be found. Note that the characteristic velocities $\lambda^{(m)}$ might be degenerate. If they were all different from each other, the system would be called totally hyperbolic.

It is instructive to use the left eigenvectors for an alternative formulation of the set of equations \eqref{eq:hyperbolic}. In fact, they can be used to find so-called Riemann invariant or characteristic variables  \cite{2013rehy.book.....R}.  To this end we start from \eqref{eq:hyperbolic} in the form
\begin{equation}
 \partial_0 \Phi_j + \left(A^{-1} B\right)_{jk} \partial_1 \Phi_k + (A^{-1} C)_j = 0.
\end{equation}
Contracting with the left eigenvectors leads to
\begin{equation}
w^{(m)}_j \partial_0 \Phi_j + \lambda^{(m)} w_j^{(m)} \partial_1 \Phi_j + w_j^{(m)}(A^{-1} C)_j = 0.
\label{eq:phileftEV}
\end{equation}
If one now introduces new variables $J^{(m)}$ such that
\begin{equation}
\label{eq:Riemann_equation}
d J^{(m)} = w_j^{(m)} d \Phi_j,
\end{equation}
the differential equation \eqref{eq:phileftEV} becomes 
\begin{equation}
\partial_0 J^{(m)} + \lambda^{(m)} \, \partial_1 J^{(m)} + w_j^{(m)} (A^{-1}C)_j = 0.
\label{eq:characteristicForm}
\end{equation}
Interestingly, we have now obtained a set of equations which is formulated in terms of derivatives along the characteristic curves labeled by the index $m$ with corresponding velocities $\lambda^{(m)}$. Without the inhomogeneous term $\sim C$, the variables $J^{(m)}$ would actually be conserved along the characteristic curves $dx^1/d x^0 = \lambda^{(m)}$. This explains the name Riemann invariants or characteristic variables. The inhomogeneous terms lead to a modified behavior which typically results in an additional damping.

Note that the causality structure of the system of partial differential equations is particularly transparent in the characteristic form \eqref{eq:characteristicForm}. In each infinitesimal time step, information encoded in the spatial dependence of the variables $J^{(m)}$ is transported with the characteristic velocity $\lambda^{(m)}$. 

The initial conditions must be given on a Cauchy curve (or a Cauchy surface for more than one space dimension) which is free and therefore non-parallel to any of the characteristic curves. The Cauchy problem is then well posed and has (locally) a unique solution. Colloquially speaking, the initial conditions are specified on a Cauchy curve, while they are propagated along the characteristic curves. 

\subsection{Causality}

The notion of causality can be made more precise for small (linear) perturbations around a given solution of the set of equations \eqref{eq:hyperbolic}. For a given space-time point $P=(x^0,x^1)$ one can show that small changes in the initial conditions outside of a certain region in the past of $P$ cannot change the solution at the point $P$. The region in the past of $P$ where small changes in the initial conditions can influence the solution at $P$ is called the domain of dependence $\Gamma_d$. In a similar way, small changes at $P$ can only affect a certain region in the future of $P$ which is called domain of influence $\Gamma_i$. The domain of dependence and domain of influence generalize the concepts of a past and future light cone familiar from electromagnetism to the present situation of non-linear (but quasi-linear) partial differential equations. 

The domain of dependence $\Gamma_d$ and domain of influence $\Gamma_i$ are bounded by the characteristic curves with minimal and maximal characteristic velocity. This illustrates again the important role of the characteristic curves for the causal structure. We discuss this further in section \ref{sec:Domainofdependenceinfluence} and in figure \ref{fig:lightcone} we illustrate the regions $\Gamma_d$ and $\Gamma_i$ for a given point in the space-time history of a heavy ion collision. In appendix \ref{appC} we recall briefly the derivation of certain inequalities (so-called energy inequalities) which allow to prove the above statements about the domain of dependence and domain of influence for linear perturbations. 

If the characteristic velocities are bounded in magnitude $|\lambda^{(m)}| \leq v_\text{max}$, the domain of dependence and domain of influence of a certain point $P$ are bound to lie in the interior of cones with velocity $v_\text{max}$, similar to light cones. As long as there is some finite maximal velocity $v_\text{max}$, one can in principle see this as a causal structure, similar to the causal structure of special (and general) relativity with $v_\text{max}$ replacing the velocity of light $c$.

In principle, if one would study a relativistic fluid in the absence of any other physical phenomena such as electromagnetism or gravity, it might be acceptable to have a causal structure with a maximal characteristic velocity $v_\text{max}$ larger than the speed of light.
Note, however that also the fluid velocity could become as large as $v_\text{max}$.
However, in practice we are not interested in such an isolated situation but want to study a QCD fluid that interacts both with electromagnetic fields and (at least in principle) gravitational fields. Moreover,  on a more microscopic level this fluid is governed by the laws of QCD as a quantum field theory, and the latter certainly imply an upper bound on the maximal velocity of signal propagation being given by the velocity of light, in agreement with the theory of relativity. It is for these physics reasons, that we demand for an effective (macroscopic) theory of a relativistic QCD fluid to follow the causality principle of special (and general) relativity according to which the maximal velocity of signal propagation, and therefore the maximal characteristic velocity, must be bounded from above by the velocity of light,
\begin{equation}
|\lambda^{(m)}| \leq c = 1.
\end{equation}
The last equality holds in our system of natural units.

\section{Fluid dynamic equations of motion}
\label{sec2}

\subsection{Hyperbolic relativistic fluid equations to second order}

The equations of motion for a relativistic fluid are the ones of energy-momentum conservation plus possibly additional conservation laws (such as for baryon number) and constitutive relations. We consider here a fluid where baryon number and other conserved quantities can be dropped and the only relevant conservation law is for energy and momentum,
\begin{equation}
\nabla_\mu T^{\mu}_{\;\;\nu} =0.
\end{equation}
The energy-momentum tensor can be decomposed as
\begin{equation}
T^{\mu}_{\;\;\nu} = \epsilon u^\mu u_\nu + (p+\pi_\text{bulk}) \Delta^{\mu}_{\;\;\nu} + \pi^{\mu}_{\;\;\nu},
\end{equation}
where the fluid velocity $u^\mu$ is defined in the Landau frame as the (unique) time-like eigenvector of $T^\mu_{\;\;\nu}$ and the energy density $\epsilon$ is the corresponding eigenvalue. 

The pressure $p$ is related to $\epsilon$ by the same relation as in thermal equilibrium, the thermodynamic equation of state $p=p(\epsilon)$ while $\pi_\text{bulk}$ is the bulk viscous pressure. The projector orthogonal to the fluid velocity is given by $\Delta^\mu_{\;\;\nu} =u^\mu u_\nu +  \delta^\mu_\nu$. Finally, the shear stress is symmetric $\pi^{\mu\nu}=\pi^{\nu\mu}$, traceless $\pi^\mu_{\;\;\mu}=0$, and transverse to the fluid velocity, $u_\mu \pi^{\mu}_{\;\;\nu}=0$. The fluid velocity itself is normalized to $u^\mu u_\mu=-1$.

The conservation law leads to the following evolution equations for the energy density and fluid velocity respectively, 
\begin{equation}
\begin{split}
\label{eq:energy_momentum_conservation}
u^\mu \partial_\mu \epsilon + \left( \epsilon + p +\pi_\text{bulk} \right) \nabla_\mu u^\mu + \pi^{\mu}_{\;\;\nu} \nabla_\mu u^\nu = & 0, \\
(\epsilon + p + \pi_\text{bulk}) u^\nu \nabla_\nu u^\mu + \Delta^{\mu\nu} \partial_\nu (p+\pi_\text{bulk}) + \Delta^{\mu\nu} \nabla_\rho \pi^{\rho}_{\;\;\nu} = & 0.
\end{split}
\end{equation}
These equations must be supplemented by constitutive relations for the shear stress $\pi^{\mu}_{\;\;\nu}$ and bulk viscous pressure $\pi_\text{bulk}$, either in the form of constraint equations or as dynamical laws. 

In what follows we shall work in the framework of a gradient expansion and include terms that are formally up to second order in gradients of temperature and fluid velocity. This setup corresponds to an expansion around local equilibrium configurations where the zeroth order corresponds to a locally equilibrated state characterized by temperature $T(x)$ and fluid velocity $u^\mu(x)$. One can also do the organization in terms of appropriately defined Knudsen and Reynolds numbers. The Knudsen number Kn is generically defined as a ratio between a microscopic scale, such as the mean free path, and a macroscopic scale of the fluid, such as the length over which the fluid velocity changes. The Reynolds number corresponds typically to the ratio of this macroscopic length scale to the dissipation scale, i.\ e.\ the length scale of perturbations that are efficiently damped by viscosity effects. In the present situation, it is convenient to define the {\it inverse} Reynolds number Re$^{-1}$ as the ratio of dissipative fields such as $\pi^{\mu\nu}$ and corresponding equilibrium fields such as pressure or enthalpy density.

The most general equation for the shear stress $\pi^{\mu}_{\;\;\nu}$ and bulk viscous pressure $\pi_{\text{bulk}}$  at second order in Knudsen number Kn and in inverse Reynolds number Re$^{-1}$ have been obtained in ref.\ \cite{Denicol:2012cn}. We are here interested in situations without a conserved particle number current. In this case, the evolution equation for the shear stress becomes

\begin{equation}
\begin{split}
\label{eq:shear_tensor_conservation_DNMR}
 P^{\mu\;\,\rho}_{\;\;\nu\;\,\sigma} \left[
 \tau_\text{shear} \left( u^\lambda \nabla_\lambda \pi^{\sigma}_{\;\;\rho} - 2  \pi^{\sigma\lambda}\omega_{\rho\lambda} \right) 
 + 2 \eta \nabla_\rho u^\sigma  
 -\varphi_7\,  \pi^\lambda_{\;\;\rho} \pi^{\sigma}_{\;\;\lambda} 
+\tau_{\pi \pi }\, \pi^{\sigma}_{\;\;\lambda} \sigma^{ \lambda}_{\;\;\rho} -\lambda_{\pi \Pi }\, \pi_{\text{bulk}} \nabla_\rho u^\sigma 
 \right] 
\\
+ \pi^{\mu}_{\;\;\nu} \left[ 1+ \delta_{\pi\pi} \nabla_\rho u^\rho
-\varphi_6\, \pi_{\text{bulk}} \right]= 0.
\end{split}
\end{equation}
We have used here the projector to the symmetric, transverse and trace-less part of a tensor,
\begin{equation}
P^{\mu\nu}_{\;\;\;\rho\sigma}=
\frac{1}{2}\Delta^{\mu}_{\;\;\rho}\Delta^\nu_{\;\;\sigma}+\frac{1}{2}\Delta^{\mu}_{\;\;\sigma}\Delta^\nu_{\;\;\rho}-\frac13 \Delta^{\mu\nu}\Delta_{\rho\sigma}.
\end{equation}
We also use the abbreviations
\begin{equation}
\sigma_{\mu\nu} = P_{\mu\nu}^{\;\;\;\rho\sigma} \nabla_\rho u_\sigma, \quad\quad \omega_{\mu\nu} = \frac{1}{2}\left( \nabla_\mu u_\nu - \nabla_\nu u_\mu \right)= \frac{1}{2}\left( \partial_\mu u_\nu - \partial_\nu u_\mu \right).
\end{equation}
Similarly for $\pi_{\mathrm{bulk}}$ one finds the evolution equation

\begin{equation}
\label{eq:bulk_pressure_conservation_DNMR}
\tau_\text{bulk} \, u^\mu \partial_\mu \, \pi_\text{bulk} + \pi_\text{bulk} + \zeta \nabla_\mu u^\mu  +\delta_{\Pi\Pi} \pi_\text{bulk}\nabla_\mu u^\mu-\varphi_1 \pi_\text{bulk}^2-\lambda_{\Pi \pi}\pi^{\mu\nu} \nabla_\mu u_\nu-\varphi_3 \pi^{\mu}_{\;\;\nu}\pi^{\nu}_{\;\;\mu}= 0.
\end{equation}

The most important term in \eqref{eq:shear_tensor_conservation_DNMR} is the one proportional to shear viscosity $\eta$ and to first order in gradients one would obtain the Navier-Stokes result $\pi^{\mu}_{\;\;\nu}=-2\eta \sigma^{\mu}_{\;\;\nu}$. Similarly, \eqref{eq:bulk_pressure_conservation_DNMR} would give to first order in gradients, the bulk viscous pressure $\pi_\text{bulk} = - \zeta \nabla_\rho u^\rho$.
At second order in gradients, in particular the relaxation times $\tau_\text{shear}$ and $\tau_\text{bulk}$ come in and eqs.\ \eqref{eq:shear_tensor_conservation_DNMR} and \eqref{eq:bulk_pressure_conservation_DNMR} mainly describe the dynamical relaxation of $\pi^{\mu}_{\;\;\nu}$ and $\pi_\text{bulk}$ towards their Navier-Stokes values. 

There are also additional, non-linear terms of second order with additional transport coefficients. More specific, the coefficients 
 $\tau_{\pi\pi}$, $\delta_{\pi\pi}$, $\lambda_{\pi\Pi}$, $\delta_{\Pi\Pi}$ and $\lambda_{\Pi\pi}$ are formally of order $\mathcal{O}(\text{Kn}\;\text{Re}^{-1})$; while 
$\varphi_7$, $\varphi_6$, $\varphi_1$ and $\varphi_3$  are formally of order $\mathcal{O}(\text{Re}^{-2})$. All these terms can be understood as non-linear modifications of how $\pi^\mu_{\;\;\nu}$ and $\pi_\text{bulk}$ relax to their asymptotic Navier-Stokes or equilibrium values. The terms of order $\mathcal{O}(\text{Kn}\;\text{Re}^{-1})$ contain one space- or time derivative acting on the dynamical fields, which may be taken to be energy density $\epsilon$ or temperature $T$, three independent components of fluid velocity $u^\mu$, five independent components of the shear stress $\pi^\mu_{\;\;\nu}$ and the bulk viscous pressure $\pi_\text{bulk}$. The terms of order $\mathcal{O}(\text{Re}^{-2})$ contain actually no derivatives.


Note that we have dropped in \eqref{eq:shear_tensor_conservation_DNMR} contributions of order $\mathcal{O}(\text{Kn}^{2})$,
that correspond to non-linear terms in the transverse gradients, like for example $(\nabla_\rho u^\rho)^2$ or $\omega^{\mu}_{\phantom{\mu}\lambda} \omega^{\lambda\nu}$. Such terms are non-linear  in derivatives of the fluid velocity and temperature and in principle appear naturally in a gradient expansion scheme. Corresponding  transport coefficients have been computed in ref. \cite{Moore:2010bu,Moore:2012tc}. However the problem with these terms is that they would transform the set of hyperbolic equations into a parabolic or mixed set of partial  differential equations which would in general be neither causal (in the relativistic sense) nor stable with respect to small perturbations. As they stand, eqs.\ \eqref{eq:shear_tensor_conservation_DNMR} and \eqref{eq:bulk_pressure_conservation_DNMR} constitute the most general set of equations for the evolution of shear stress and bulk viscous pressure at second order in the formal gradient expansion around local equilibrium but with only first order derivatives acting on the extended fluid fields $\epsilon$, $u^\mu$, $\pi^\mu_{\;\;\nu}$ and $\pi_\text{bulk}$. These equations are hyperbolic and have therefore a chance to be causal and stable with respect to linear perturbations. 

\subsection{Coordinate system}
\label{sec:CoordinateSystem}

We will now consider a more concrete situation of a relativistic nuclear collisions at high energies with a particular set of symmetries. We discuss here first coordinate systems that are particularly useful for our purpose. 

Cartesian coordinates may be chosen such that the $z$-axis agrees with the beam axis, that the collision of the strongly Lorentz contracted nuclei takes place at $t=z=0$, and that the center of the overlap region is at the origin of the transverse plane $x=y=0$. We follow Bjorken in assuming that the dynamics is invariant under longitudinal boosts. This should be a good approximation at large collision energy and in the central longitudinal region.

It is therefore convenient to introduce the proper time $\tau=\sqrt{t^2-z^2}$ and longitudinal rapidity $\eta$ such that $t=\tau \cosh(\eta)$ and $z=\tau \sinh(\eta)$. Moreover, we introduce polar coordinates in the transverse plane such that $x=r \cos(\phi)$ and $y=r \sin(\phi)$. In these coordinates, the Minkowski space metric becomes
\begin{equation}
ds^2=-d\tau^2 + dr^2 + r^2 d\phi^2 + \tau^2 d\eta^2.
\label{eq:metricBjorkenAzimuthal}
\end{equation}

Below we will mainly concentrate on central collisions where the initial conditions are invariant under translations in rapidity, $\eta\to \eta+\Delta\eta$, and the azimuthal angle, $\phi\to \phi  + \Delta\phi$, corresponding to Bjorken boosts and azimuthal rotations, respectively. By symmetry reasons, this will then also be the case at later times. The dynamics is therefore effectively reduced from 3+1 space and time dimensions to the 1+1 dimensional subspace of proper time $\tau$ and transverse radius $r$.

To discuss the causal  structure of the transverse expansion dynamics it will be convenient to introduce another parametrization that is related to $\tau$ and $r$ by a conformal mapping similar to a Penrose diagram \cite{Penrose:1964ge}. We write
\begin{equation}
r-\tau = h(\rho-\sigma), \quad\quad\quad r+\tau = h(\rho+\sigma),
\label{eq:transformh}
\end{equation}
where the new coordinates $\sigma$ and $\rho$ replace $\tau$ and $r$. The monotonic function $h(x)$ may be chosen to be a map from the interval $(-1,1)$ to $(-\infty, \infty)$ with the following properties
\begin{equation}
h'(x)>0, \quad \quad\quad h(0)=0, \quad\quad\quad \lim_{x\to 1}h(x)=+\infty,\quad\quad\quad h(x)=-h(-x).
\end{equation}
As an example take
\begin{equation}
\label{eq:hfunction}
h(x) = \text{sign}(x) R \left[ \text{arctanh}(|x|^\alpha) \right]^{1/\alpha},
\end{equation}
with some length $R>0$ and exponent $\alpha>0$.
 In this case, the infinite sector $0\leq \tau, r < \infty$ is parametrized by a finite region in terms of $\sigma$ and $\rho$ namely $0\leq \sigma <1$ and $0\leq \rho < 1-\sigma$. This coordinate system may also be useful for a numerical treatment because only a finite coordinate region must be considered.

\begin{figure}
 \label{fig:coordinates_system}
\centering 
\subfigure[]{\label{fig:coordinates_system-a}
\includegraphics[width=0.45\textwidth]{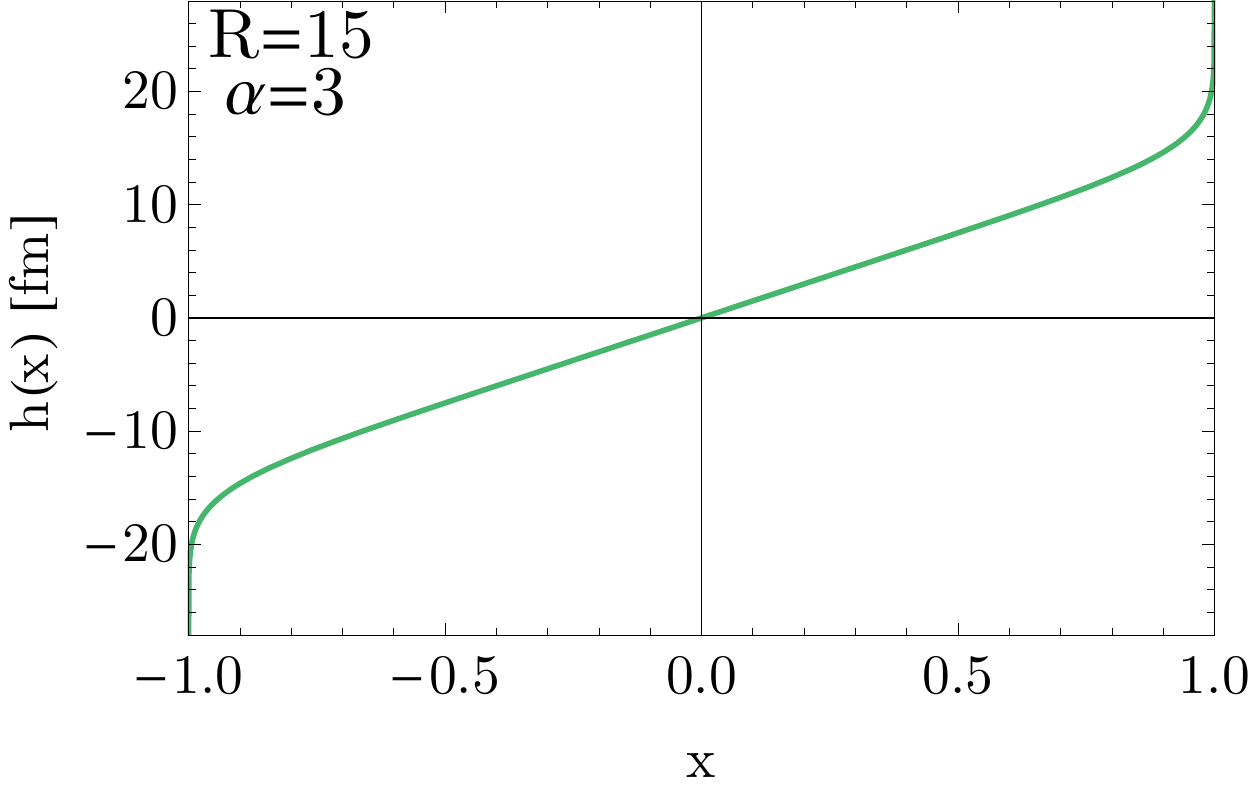}
}
\subfigure[]{\label{fig:coordinates_system-b}
\includegraphics[width=0.45\textwidth]{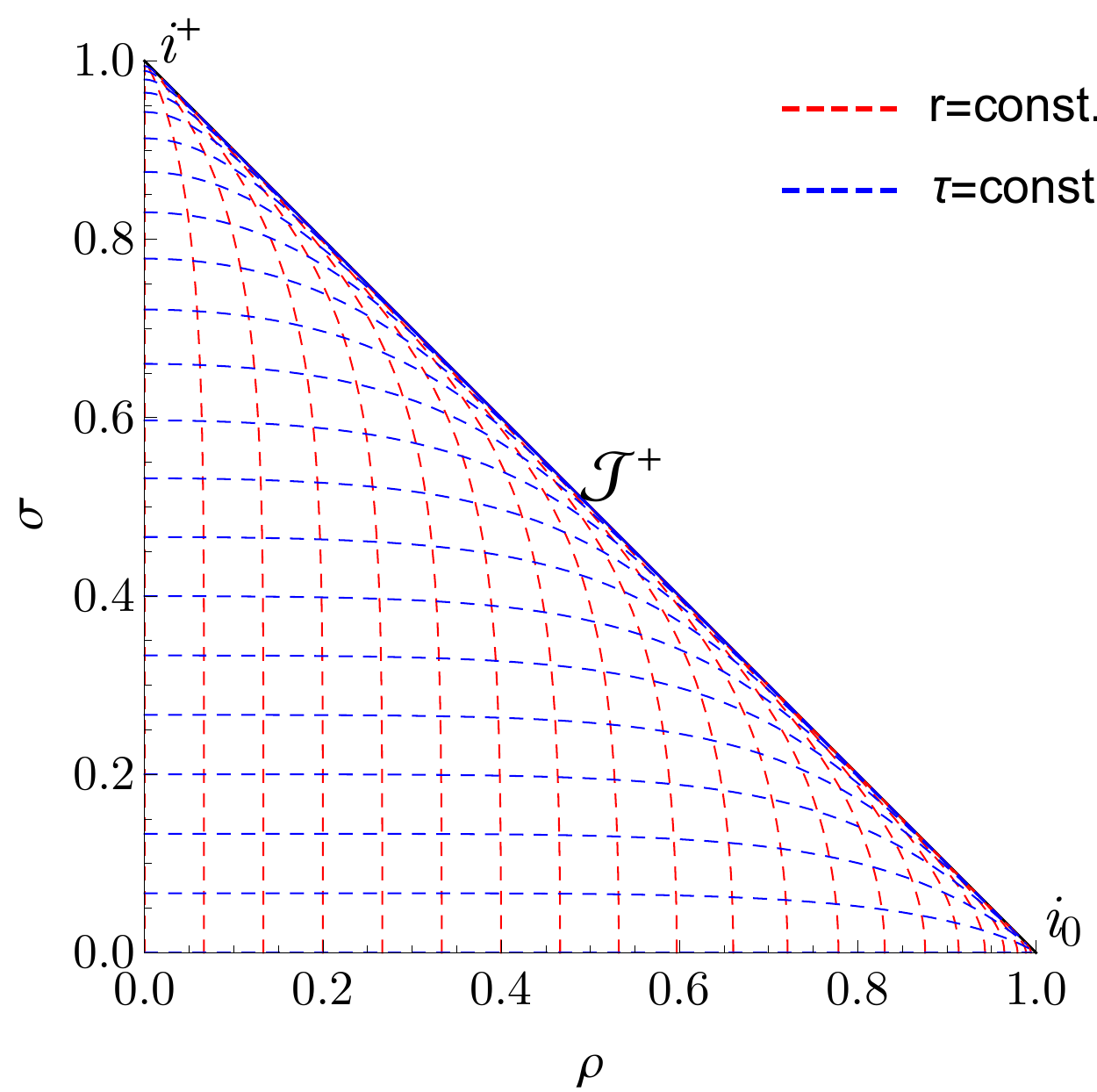}
}
\caption{
 (a) Function $h(x)$ used to define the coordinates $\sigma$ and $\rho$ in eq. \eqref{eq:transformh}. We use the form  \eqref{eq:hfunction} with $R=15 \text{ fm}$ and $\alpha=3$.
(b) Coordinate mesh of constant Bjorken time $\tau$ and transverse radius $r$ as a function of the coordinates $0\le\sigma,\, \rho\le1$.
The connection between $\tau,r$ and $\sigma,\rho$ is a conformal map in the two-dimensional space, similar to a Penrose diagram. The infinite half-plane  $0\le \tau,r \le \infty $ is covered by a finite region in terms of $\sigma $ and $\rho$.
The point $i_0$ at $\sigma =0$ and $\rho =1$ corresponds to spatial infinity $r\to \infty$ at fixed $\tau$.
Similarly $i^+$ at $\sigma=1$ and $\rho=0$ corresponds to timelike infinity $\tau \to\infty$ with fixed finite $r $.
The connecting line $\mathcal{J}^+$ corresponds to lightlike infinity $\tau\to \infty$ and $r \to \infty$ at fixed ratio $\tau /r $.
}
\end{figure}

In figure \ref{fig:coordinates_system-a} we display the function $h(x)$ in eq.\ \eqref{eq:hfunction} and in fig. \ref{fig:coordinates_system-b} we show the coordinate lines of constant $r$ and $\tau$ as a function of $\rho$ and $\sigma$.
Note that the point $\rho=1$, $\sigma=0$ corresponds to spatial infinity $r\to \infty$ with fixed time $\tau$ and is labeled by $i_0$ in fig. \ref{fig:coordinates_system-b}.
Similarly, the point $\sigma=1$, $\rho=0$ corresponds to $\tau\to \infty$ with fixed $r$, i.e. timelike infinity and is labeled by $i^+$ in figure \ref{fig:coordinates_system-b}. 
Finally, the line $\sigma=1-\rho$ corresponds to $\tau \to\infty $, $r\to \infty $ with fixed ratio $\tau/r$. This corresponds to lightlike infinity and is labeled by $\mathcal{J^+}$ in figure \ref{fig:coordinates_system-b}.
The Minkowski space metric becomes
\begin{equation}
ds^2 = h^\prime(\rho-\sigma) h^\prime(\rho+\sigma) [-d\sigma^2+d\rho^2] + \frac{1}{4}[h(\rho-\sigma)+h(\rho+\sigma)]^2 d\phi^2 + \frac{1}{4}[-h(\rho-\sigma)+h(\rho+\sigma)]^2 d\eta^2.
\label{eq:metricConformalBjorken}
\end{equation}
Note that for $d\phi=d\eta=0$ the metric \eqref{eq:metricConformalBjorken} is indeed related to the metric \eqref{eq:metricBjorkenAzimuthal} by a conformal transformation, in particular light rays $d\tau=\pm dr$ preserve their form and become $d\sigma=\pm d\rho$. This feature is particularly useful to investigate the issue of causality. We note as a side remark that \eqref{eq:transformh} is not a conformal transformation in the full four-dimensional sense.

Note also that the metric \eqref{eq:metricBjorkenAzimuthal} and \eqref{eq:metricConformalBjorken} are special cases of the ansatz
\begin{equation}
g_{\mu\nu} = \text{diag}(-g_{11}, g_{11}, g_{22}, g_{33}),
\end{equation}
with $g_{11}$, $g_{22}$ and $g_{33}$ being functions of the time coordinate $x^0$ and radial coordinate $x^1$. In the following it will often be convenient to work with this general case and restrict to the special cases of Bjorken coordinates $\tau$, $r$, $\phi$ and $\eta$ or the conformally related coordinates $\sigma$, $\rho$, $\phi$ and $\eta$ at other places. We collect useful formulas for these coordinate choices, such as Christoffel symbols, projection operators etc.\ in appendix \ref{appA}.

\subsection{Evolution equations for radial expansion}

Assuming longitudinal boost invariance as well as azimuthal rotation symmetry, the fluid dynamic equations of motion become partial differential equations with one time coordinate $x^0$ and one radial coordinate $x^1$. For the set of equations \eqref{eq:energy_momentum_conservation}, \eqref{eq:shear_tensor_conservation_DNMR} and \eqref{eq:bulk_pressure_conservation_DNMR}, the equations are generically of hyperbolic type, and can be formulated as a first order system of equations for conveniently chosen independent fields. For our numerical treatment we will take as independent fields 
\begin{itemize}
\item the radial fluid rapidity $\chi$ in terms of which the properly normalized fluid velocity is given by
\begin{equation*}
u^\mu = \left( \frac{\text{cosh}(\chi)}{\sqrt{g_{11}}}, \frac{\text{sinh}(\chi)}{\sqrt{g_{11}}}, 0, 0 \right),
\end{equation*}
\item the temperature $T$ or actually its logarithm $\ln (T)$,
\item two independent components of shear stress normalized by the enthalpy density $\tilde \pi^\phi_\phi=\pi^3_{\;\;3}/(\epsilon+p)$ and $\tilde \pi^\eta_\eta=\pi^4_{\;\;4}/(\epsilon+p)$ to which the other components are related by eq.\ \eqref{eq:parametrizationShearStress},
\item and the bulk viscous pressure normalized by enthalpy density $\tilde \pi_\text{bulk}=\pi_\text{bulk}/(\epsilon+p)$.
\end{itemize}
All these fluid variables are dimensionless and unconstrained fields, for which evolution equations follow from the set of equations \eqref{eq:energy_momentum_conservation}, \eqref{eq:shear_tensor_conservation_DNMR} and \eqref{eq:bulk_pressure_conservation_DNMR}.

These equations can be formulated as a quasi-linear differential equation
\begin{equation}
A_{ij} \partial_0 \Phi_j + B_{ij} \partial_1 \Phi_j + C_i = 0 ,
\label{eq:quasilinearmatix}
\end{equation}
with the combined field $\Phi=(\chi, \ln T, \tilde\pi^\phi_\phi, \tilde\pi^\eta_\eta, \tilde\pi_\text{bulk})$ and coefficients $A_{ij}$, $B_{ij}$ and $C_i$ that are functions of $\Phi$ but not of its derivatives. More concrete, we present the full set of equations of motion in appendix \ref{app:EOM} and we obtain in particular the matrix associated to the time derivative
\begin{equation}
A = \begin{pmatrix} d_1 \cosh(\chi) &
d_2 \sinh(\chi ) & -\sinh(\chi ) & -\sinh(\chi ) & \sinh(\chi ) \\
 d_1 \sinh(\chi ) &
   \frac{1}{c_s^2} \cosh(\chi ) & 0 & 0 & 0 \\
 d_3 \sinh(\chi ) & \tau_\text{shear} \tilde\pi^\phi_\phi (1+\frac{1}{c_s^2}) \cosh(\chi) & \tau_\text{shear} \cosh(\chi ) & 0 & 0 \\
 d_4  \sinh(\chi ) & \tau_\text{shear} \tilde\pi^\eta_\eta(1+\frac{1}{c_s^2}) \cosh(\chi) & 0 & \tau_\text{shear} \cosh(\chi ) & 0 \\
 d_5  \sinh(\chi ) & \tau_\text{bulk} \tilde\pi_\text{bulk} (1+\frac{1}{c_s^2}) \cosh(\chi) & 0 & 0 & \tau_\text{bulk} \cosh(\chi ) \end{pmatrix},
\end{equation}
and the one associated to the radial derivative
\begin{equation}
B = \begin{pmatrix} d_1 \sinh(\chi) &
 d_2 \cosh(\chi ) & -\cosh(\chi ) & -\cosh(\chi ) & \cosh(\chi ) \\
d_1 \cosh(\chi ) &
   \frac{1}{c_s^2} \sinh(\chi ) & 0 & 0 & 0 \\
 d_3 \cosh(\chi ) & \tau_\text{shear} \tilde\pi^\phi_\phi (1+\frac{1}{c_s^2}) \sinh(\chi) & \tau_\text{shear} \sinh(\chi ) & 0 & 0 \\
 d_4 \cosh(\chi ) & \tau_\text{shear} \tilde\pi^\eta_\eta(1+\frac{1}{c_s^2}) \sinh(\chi) & 0 & \tau_\text{shear} \sinh(\chi ) & 0 \\
 d_5 \cosh(\chi ) & \tau_\text{bulk} \tilde\pi_\text{bulk} (1+\frac{1}{c_s^2}) \sinh(\chi) & 0 & 0 & \tau_\text{bulk} \sinh(\chi ) \end{pmatrix}.
\end{equation}
We have used here the abbreviations

\begin{equation}
\label{eq:abbreviations}
\begin{split}
d_1 & = 1 + \tilde \pi_\text{bulk}-\tilde\pi^\phi_\phi-\tilde\pi^\eta_\eta, \\
d_2 & = 1 + \left(1+\frac{1}{c_s^2}\right) \left(\tilde\pi_\text{bulk}-\tilde\pi^\phi_\phi-\tilde\pi^\eta_\eta \right),\\
d_3 & =-\frac{1}{3}\left(\frac{2\eta}{\epsilon+p} +\tau_{\pi\pi} \tilde\pi^{\phi}_\phi-\lambda_{\pi\Pi}\tilde\pi_{\text{bulk}} \right)
+\delta_{\pi\pi} \tilde\pi^{\phi}_\phi, \\
d_4 & =-\frac{1}{3} \left(\frac{2\eta}{\epsilon+p} +\tau_{\pi\pi} \tilde\pi^{\eta}_\eta-\lambda_{\pi\Pi}\tilde\pi_\text{bulk} \right)
+  \delta_{\pi\pi} \tilde\pi^{\eta}_\eta, \\
d_5 & = \frac{\zeta}{\epsilon+p}+\delta_{\Pi\Pi} \tilde\pi_\text{bulk} +\lambda_{\Pi\pi} \left(\tilde\pi^\eta_\eta+\tilde\pi^\phi_\phi \right).
\end{split}
\end{equation}

The ``source terms'' $C_i$ are somewhat lengthy and we do not display them explicitly. However, they can easily be read off from the equations in appendix \ref{app:EOM}.
Together with initial values provided on an appropriate Cauchy surface (for example  $\tau =\tau_0$), the differential equation \eqref{eq:quasilinearmatix} specifies the solution of the fluid equations completely.


\subsection{Characteristics of the differential equations}

As we have reviewed in section \ref{eq:HyperbolicPDEs}, quasi-linear systems of differential equations such as \eqref{eq:quasilinearmatix} can be characterized in terms of their characteristics. The characteristic curves can be seen as the lines along which information is transported, for example small perturbations, discontinuities, defects or shocks. The system is causal in the relativistic sense precisely if the characteristic velocities are smaller than (or, as a limit, equal to) the velocity of light.

More specific, the solution of the differential equation \eqref{eq:quasilinearmatix} at a particular space-time point $x$ has a domain of dependence in the past of that point bounded by the characteristics  with largest and smallest characteristic velocities \cite{1953mmp..book.....C}. This will be illustrated in more detail below.

In order to find the characteristics of eq.\ \eqref{eq:quasilinearmatix} we need to solve the eigenvalue problem
\begin{equation}
w_i^{(n)} B_{ij}  = \lambda^{(n)} \,w_i^{(n)} \, A_{ij} ,
\label{eq:evproblemcharacteristics}
\end{equation}
where the eigenvalues $\lambda^{(n)}$ correspond to the characteristic velocities and $w_i^{(n)}$ are the corresponding left eigenvectors. 
The eigenvalues $\lambda^{(n)}$ corresponding to the characteristic velocities follow from the condition 
\begin{equation}
\det\left( B-\lambda^{(n)} A \right)=0.
\end{equation}
A direct calculation shows that they are given by 
\begin{equation}
\lambda^{(1)} = \frac{v+\tilde{c}}{1+\tilde{c} v}, \quad\quad\quad \lambda^{(2)} = \frac{v-\tilde{c}}{1-\tilde{c} v}, \quad\quad\quad \lambda^{(3)} = \lambda^{(4)} = \lambda^{(5)}  = v,
\label{eq:eigenvaluesCharacteristics}
\end{equation}
where $v=\tanh(\chi)$ is the fluid velocity and $\tilde{c}$ is a modified sound velocity which we discuss below. Note that the characteristic velocities $\lambda^{(1)}$ and $\lambda^{(2)}$ correspond to ``relativistic sums'' of the fluid velocity $v$ and the modified speed of sound $\tilde{c}$. Causality $|\lambda^{(n)}|\leq 1$ is guaranteed when $|v|\leq 1$ and $|\tilde{c}|\leq 1$.

Physically, we expect that the two characteristics with velocities $\lambda^{(1)}$ and $\lambda^{(2)}$ describe generalized sound propagation including viscous as well as non-linear effects. In contrast, the characteristics with velocities $\lambda^{(3)}$, $\lambda^{(4)}$ and $\lambda^{(5)}$ should be understood as non-linear generalizations of diffusive modes for which information is propagated along the fluid flow lines. 

Let us now discuss the modified sound velocity $\tilde{c}$. It can be written as
\begin{equation}
\label{eq:modified_cs}
\tilde{c}=\sqrt{c_s^2 + d},
\end{equation}
where the ideal fluid velocity of sound is determined by the thermodynamic relation
\begin{equation}
c_s^2 = \frac{\partial p}{\partial \epsilon}=\frac{\frac{\partial p}{\partial T}}{\frac{\partial \epsilon}{\partial T}},
\label{eq:ideal_cs}
\end{equation}
and the viscous and non-linear modification is parametrized by the combination

\begin{equation}
\begin{split}
\label{eq:explicit_d}
d = & \frac{
\frac{4\eta}{3\tau_\text{shear}}+ \frac{\zeta}{\tau_\text{bulk}}-
\left( \frac{\tau_{\pi\pi}}{3\tau_\text{shear}}-\frac{\delta_{\pi\pi}}{\tau_\text{shear}}+\frac{\lambda_{\Pi\pi}}{\tau_\text{bulk}}\right)\left(\pi^\phi_\phi+\pi^\eta_\eta \right)+\left(\frac{\delta_{\Pi\Pi}}{\tau_\text{bulk}}+\frac{\lambda_{\pi\Pi}}{3\tau_\text{shear}}\right)\pi_\text{bulk}
}
{\epsilon + p + \pi_\text{bulk} - \pi^\phi_\phi - \pi^\eta_\eta } \\
= & \frac{
\frac{4\eta}{3\tau_\text{shear}(\epsilon+p)}+ \frac{\zeta}{\tau_\text{bulk}(\epsilon+p)}-
\left( \frac{\tau_{\pi\pi}}{3\tau_\text{shear}}-\frac{\delta_{\pi\pi}}{\tau_\text{shear}}+\frac{\lambda_{\Pi\pi}}{\tau_\text{bulk}}\right)\left(\tilde\pi^\phi_\phi+\tilde\pi^\eta_\eta \right)+\left(\frac{\delta_{\Pi\Pi}}{\tau_\text{bulk}}+\frac{\lambda_{\pi\Pi}}{3\tau_\text{shear}}\right)\tilde \pi_\text{bulk}
}
{1 + \tilde\pi_\text{bulk} - \tilde \pi^\phi_\phi - \tilde \pi^\eta_\eta }\\
= & - \frac{d_3+d_4}{d_1 \tau_\text{shear}}+\frac{d_5}{d_1 \tau_\text{bulk}}.
\end{split}
\end{equation}
%

The second equation uses the dimensionless variables introduced in appendix \ref{appA} and the third equation uses the abbreviations \eqref{eq:abbreviations}. We note that causality requires large enough relaxation times $\tau_\text{shear}$ and $\tau_\text{bulk}$ for given shear viscosity $\eta$ and bulk viscosity $\zeta$. In fact, this was already known from the analysis of small perturbations around thermal equilibrium states by Hiscock and Lindblom \cite{Hiscock:1983zz}. In contrast to ref. \cite{Hiscock:1983zz}, our analysis applies for the general, non-linear evolution in the specific situation of azimuthally symmetric and Bjorken boost invariant heavy ion collisions. 

Note that the causality constraint formulated here is not only a constraint on thermodynamic and transport properties. At the non-linear level (in deviations from equilibrium) it involves the shear stress as parametrized in terms of $\pi^\phi_\phi$ and $\pi^\eta_\eta$ as well as the bulk viscous pressure $\pi_\text{bulk}$. One needs to check for a specific solution of the fluid equations whether the causality constraint $|\tilde{c}|\leq 1$ is satisfied, and we will do so for a typical heavy ion collision below.

A set of left eigenvectors corresponding to \eqref{eq:evproblemcharacteristics} and the eigenvalues \eqref{eq:eigenvaluesCharacteristics} is given by
\begin{equation}
\begin{split}
\label{eq:left_eigenvectors}
w^{(1)} = & \begin{pmatrix} 1, & \frac{c_s^2 d_1 + d_1 -1}{c_s^2 \tilde{c}\;d_1 }, & 
- \frac{1}{ \tilde{c}\; d_1}, & - \frac{1}{\tilde{c}\; d_1}, & \frac{1}{ \tilde{c}\; d_1} \end{pmatrix}, \\
w^{(2)} = & \begin{pmatrix} 1, & -\frac{c_s^2 d_1 + d_1 -1}{c_s^2 \tilde{c}\;d_1 }, & 
 \frac{1}{ \tilde{c}\; d_1}, &  \frac{1}{\tilde{c}\; d_1}, & -\frac{1}{ \tilde{c}\; d_1} \end{pmatrix}, \\
w^{(3)} = & \begin{pmatrix} 0, &\frac{(1+c_s^2 )\pi^\phi_\phi}{ c_s^2}-\frac{d_3}{\tau_\text{shear} d_1 c_s^2} , &1, & 0, &
0 \end{pmatrix},\\ 
w^{(4)} = & \begin{pmatrix} 0, &\frac{(1+c_s^2 )\pi^\eta_\eta}{ c_s^2}-\frac{d_4}{\tau_\text{shear} d_1 c_s^2}, & 0, & 1, & 0\end{pmatrix},\\ 
w^{(5)} = & \begin{pmatrix} 0, & \frac{(1+c_s^2 )\pi_{\text{bulk}}}{ c_s^2}-\frac{d_5}{\tau_\text{bulk} d_1 c_s^2}, & 0, & 0,&1 \end{pmatrix}.
\end{split}
\end{equation}

\section{Causality of the radial expansion}
\subsection{Causal initial condition and evolution}


 \begin{figure}[tbp]
\centering 
\subfigure[]{\label{fig:speed_of_sound-a}
\includegraphics[width=0.45\textwidth]{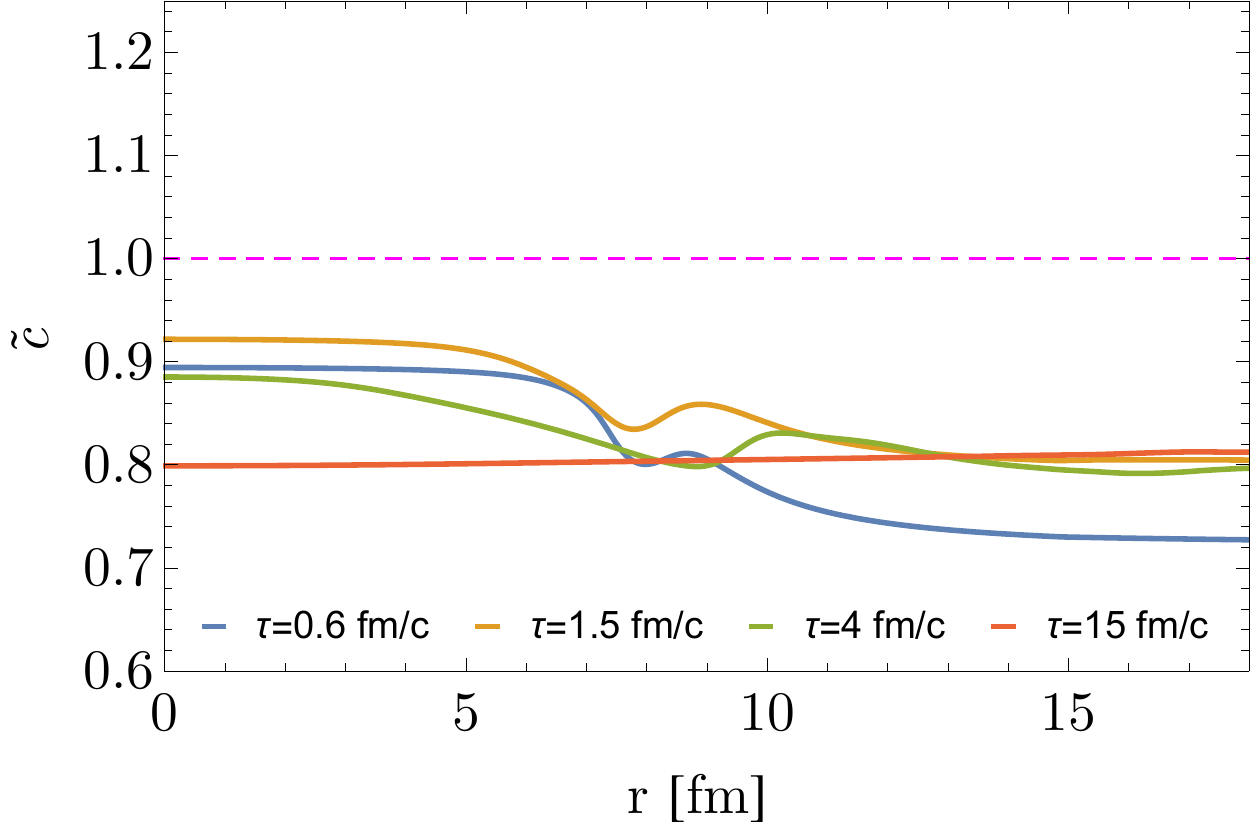}}
\subfigure[]{\label{fig:speed_of_sound-b}
\includegraphics[width=0.45\textwidth]{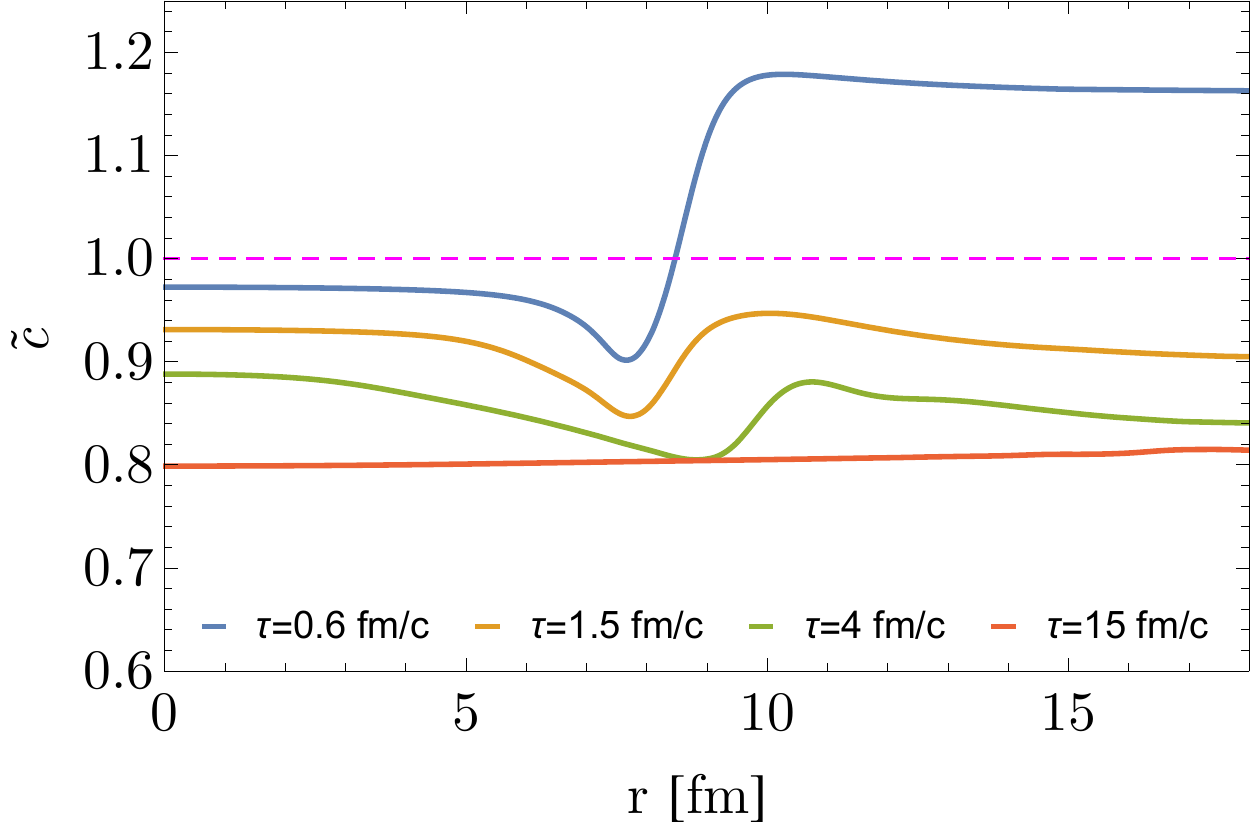}}
\\
\subfigure[]{\label{fig:speed_of_sound-c}
\includegraphics[width=0.45\textwidth]{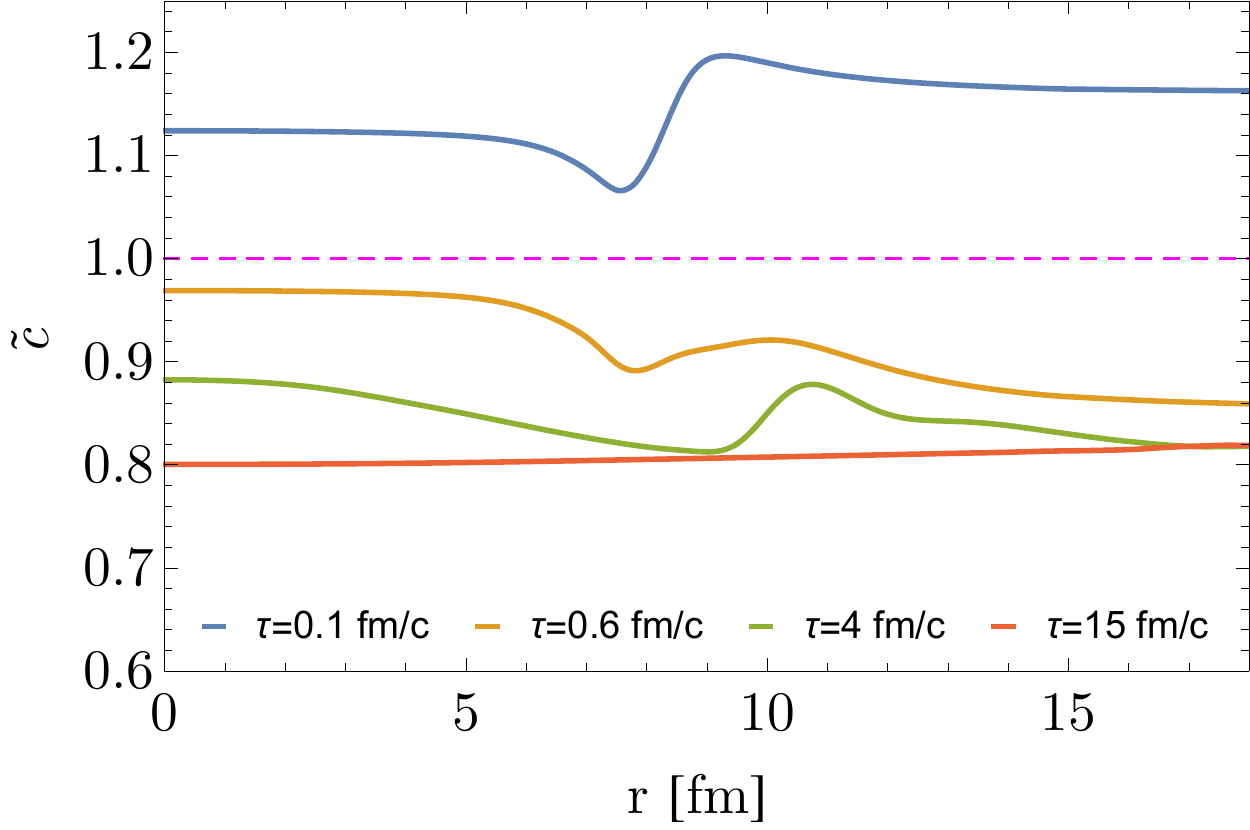}}
\subfigure[]{\label{fig:speed_of_sound-d}
\includegraphics[width=0.45\textwidth]{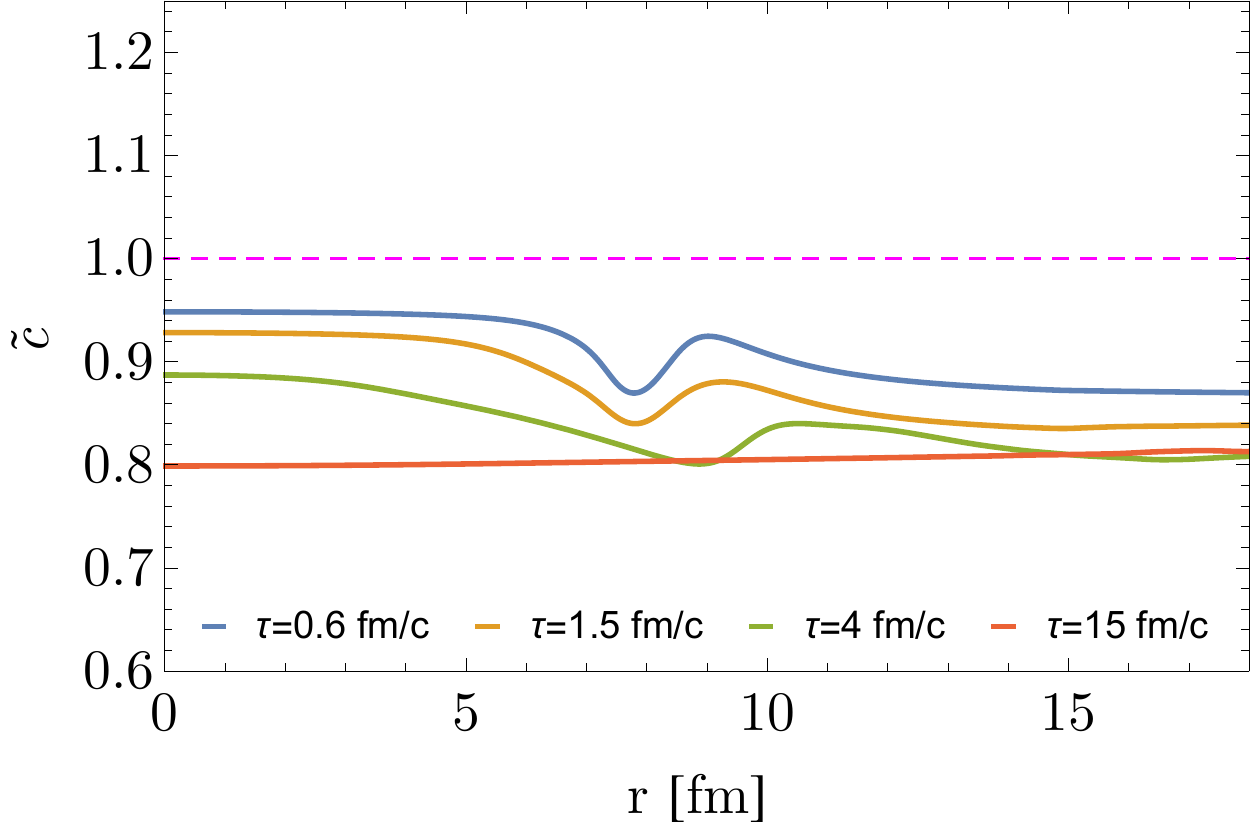}}
\caption{\label{fig:speed_of_sound}
Modified speed of sound $\tilde c$ according to \eqref{eq:modified_cs} 
including viscous and non-linear effects, as a function of a radius $r$ for different Bjorken times $\tau$.
We compare different initializations of the shear stress components. 
In (a) we have set $\tilde \pi^\phi_\phi=\tilde \pi^\eta_\eta=0$ at the initialization time $\tau_0=0.6$ fm/c. The causality constraints $|\tilde c|\le1$ is satisfied there as well as at subsequent   times.
In (b) and (c) we have chosen Navier-Stokes  initial conditions according to equation \eqref{eq:shear_stress_Navier_Stokes} at the initialization time $\tau_0=0.6$ fm/c and $\tau_0=0.1$ fm/c, respectively.
One observes that in the former case causality is violated at early times and for large radii and in the latter case for early times at all radii.
For later times one finds decreasing $\tilde c$ such that $|\tilde c| \le 1$ becomes valid.
In (d) we have used a modification of Navier-Stokes 
initial condition according to equation \eqref{eq:shear_stress_initial_condition}. The causality constraint is satisfied then at all times $\tau$ and radii $r$.
}
\end{figure}

Let us now investigate the condition for a causal evolution in more detail.
The characteristic velocities  in \eqref{eq:eigenvaluesCharacteristics} are below the speed of light and the system of first order hyperbolic equations \eqref{eq:quasilinearmatix} is accordingly causal, precisely if the modified sound velocity $\tilde{c}$ in \eqref{eq:modified_cs} is bounded by $|\tilde{c}|\le 1$. 
This in turn is a condition that involves thermodynamic quantities such as the ideal fluid velocity of sound \eqref{eq:ideal_cs}, (ratios of) transport coefficients such as $\eta /\tau_\text{shear}$ but also the components of shear stress $\pi^\eta_\eta$ and $\pi^\phi_\phi$ as well as the viscous pressure $\pi_{\text{bulk}}$.
The initialization  of the fluid evolution includes also the specification of initial values for the shear stress and bulk viscous pressure. This means that the causality bounds $|\tilde{c}|\le 1 $ is also a condition for a viable initial conditions.

To investigate how important this causality bound is in practice, we will now determine the modified sound velocity $\tilde c $ for typical initial conditions 
relevant  to high energy nuclear collisions.

For the numerical investigation we use  the QCD equation of state provided by 
\cite{Borsanyi:2016ksw}
and  we  adopt a temperature dependent  shear-viscosity \cite{Christiansen:2014ypa} calculated for Yang-Mills theory.
We neglect the bulk viscous pressure.
Among the second order transport coefficient we keep only $\tau_{\text{shear}}$ and  $\delta_{\pi\pi}$, while other second  order coefficients do not enter or are neglected. 
We have chosen the value of this transport coefficient such that 
\begin{equation}
\label{eq:transport_properties}
\tau_{\text{shear}}=\eta \frac{2(2-\text{ln}\left(2\right)) }{\epsilon+p},\quad\quad\quad\quad \delta_{\pi\pi}= \frac{4}{3}\tau_{\text{shear}},
\end{equation}
as obtained from AdS/CFT calcualtions  \cite{Baier:2007ix}. We initialize fluid dynamics on the hypersurface $\tau =\tau_0$ with vanishing 
radial fluid velocity $\chi(\tau_0)=0$ and we choose the initial temperature profile 
according to the model discussed in ref \cite{Qiu:2011hf}, with a maximal temperature in the center of the fireball of $T_{\text{max}}=0.4$ GeV.

For the initial condition of the shear stress we compare four possibilities:
\begin{itemize}
\item[(a)] Vanishing azimutal and rapidity component
$$
\tilde \pi^\phi_\phi=\tilde \pi^\eta_\eta=0,
$$
with initialization time $\tau_0=0.6$ fm/c.
\item[(b)] Navier-Stokes initial condition $\pi^{\mu}_{\nu}=-2\eta \sigma^\mu_\nu$
initialized at $\tau_0=0.6$ fm/c, which leads to
\begin{equation}
\label{eq:shear_stress_Navier_Stokes}
\sqrt{g_{11}} \tilde\pi^\phi_\phi -\frac{2\eta}{3(\epsilon+p)} \partial_0 \text{ln} (\sqrt{g_{11}}\sqrt{g_{33}}/g_{22})=0,
\quad\quad 
\sqrt{g_{11}} \tilde\pi^\eta_\eta -\frac{2\eta}{3(\epsilon+p)} \partial_0 \text{ln} (\sqrt{g_{11}}\sqrt{g_{22}}/g_{33})=0.
\end{equation}
\item[(c)] Navier-Stokes initial condition as in \eqref{eq:shear_stress_Navier_Stokes} but now initialized as $\tau_0=0.1$ fm/c.
\item[(d)] A variant of the Navier-Stokes initial condition specified by
\begin{equation}
\label{eq:shear_stress_initial_condition}
\sqrt{g_{11}} \tilde\pi^\eta_\eta -\frac{2\eta}{3(\epsilon+p)} \partial_0 \text{ln} (\sqrt{g_{11}}\sqrt{g_{22}}/g_{33})+\delta_{\pi\pi}\tilde\pi^\eta_\eta \partial_0 \text{ln}(\sqrt{g_{11}}\sqrt{g_{22}}\sqrt{g_{33}})=0,
\end{equation}
and similarly for $\tilde \pi^\phi_\phi$.
\end{itemize}  

For the four choices of initial conditions above we show the velocity parameter
$\tilde{c}$ as a function of the radius $r$ in figure \ref{fig:speed_of_sound}.
We also plot $\tilde{c}$ at later times $\tau$ as obtained from solving equations  \eqref{eq:quasilinearmatix} numerically.

One observe that the initial condition (a) leads to $ |\tilde{c}|\le1$ for all times $\tau $ and radii $r$.
Relativistic causality is indeed satisfied here. In contrast, the Navier-Stokes initial conditions (b) and (c) are not viable from a causality point of view. With the initialization time $\tau_0=0.6$ fm/c corresponding to (b), the velocity $\tilde{c}$ 
exceed the velocity of light for larger radii $r$ at early time. At later time $\tau$
the shear stress relaxes and $\tilde{c}$ decreases below the velocity of light.
If the initialization time is chosen as $\tau_0=0.1$ fm/c  corresponding  to (c),
the causality bound is actually violated at all radii $r$ for early time $\tau$.
The Navier-Stokes initial condition is therefore not viable  at early times $\tau$
in the second order approximation for fluid dynamics \eqref{eq:shear_tensor_conservation_DNMR}, althought the solution relaxes 
towards smaller value of $\tilde c $ at late time. Finally, case (d) corresponds to a generalization of Navier-Stokes initial condition 
that involves also the additional transport coefficient $\lambda_2$. As one can read off from \ref{fig:speed_of_sound-d}, when used at the initialization time $\tau_0=0.6$ fm/c, this leads to $|\tilde{c}|\le 1$ at all relevant times and radii $r$. 

The violation of the causality bound for Navier-Stokes initial condition at early time can be also be highlighted directly from equation \eqref{eq:explicit_d}.The initial conditions can be written in a simple form
$$
\tilde\pi^\phi_\phi+\tilde\pi_\eta^\eta=-\frac{2\eta }{3(\epsilon+p)\tau_0},
$$
and the  causality constraint on the modified sound velocity $|\tilde c|<1$ leads to the condition
\begin{equation}
\label{eq:tau_inequality}
\frac{\tau_0}{\tau_\text{shear}} \ge  \frac{c_s^2+\frac{\delta_{\pi\pi}}{\tau_\text{shear}}-1}{1-c_s^2-2\Gamma}\Gamma,
\end{equation}
where we have used the abbreviation
$$
\Gamma =\frac{2\eta}{3(\epsilon +p )\tau_\text{shear}}.
$$
In general, the right-hand side is a function of the initial temperature,
but 
in order to have an estimation of magnitude of the bound, we can approximate the speed of sound by $c_s^2= 1/3$, while \eqref{eq:transport_properties} leads to  $\Gamma \approx 2/9$ and for $\delta_{\pi\pi}$we can take the conformal value $ \delta_{\pi\pi} =4/3 \tau_\text{shear}$. Inserting this values into the inequality \eqref{eq:tau_inequality} we obtain
\begin{equation}
\frac{\tau_0}{\tau_\text{shear}} \ge \frac{2}{9},
\end{equation}
and restoring the units we have
\begin{equation}
\tau_0\ge 0.3 \,\,\,\frac{\text{fm}}{\text{c}}.
\end{equation}
This simple estimation explains the differences between the initial condition  
(b) and (c) in the center of the fireball, where the first is initialized at $\tau_0=0.6 \,\,\,\text{fm}/\text{c} $, consequently the causality condition is satisfied while (c) is initialized at $\tau_0=0.1 \,\,\,\text{fm}/\text{c} $ and the causality bound is violated. 
%

\subsection{Characteristic curves } 
\label{sec:CharacteristicCurves }

\begin{figure}[tbp]
\centering 
\subfigure[]{\label{fig:PenroseDiagram-a}
\includegraphics[width=0.40\textwidth]{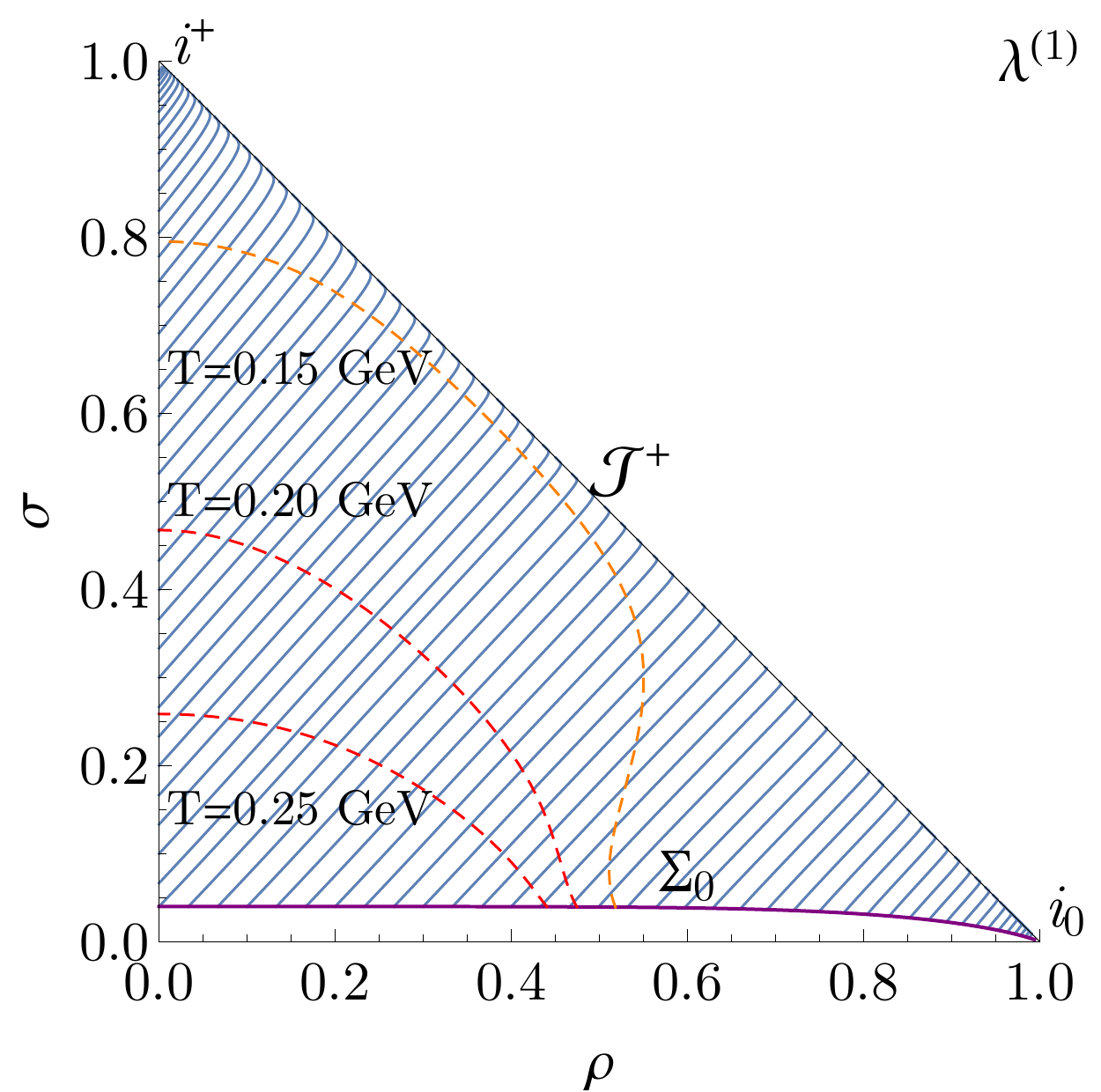}}
\subfigure[]{\label{fig:PenroseDiagram-b}
\includegraphics[width=0.40\textwidth]{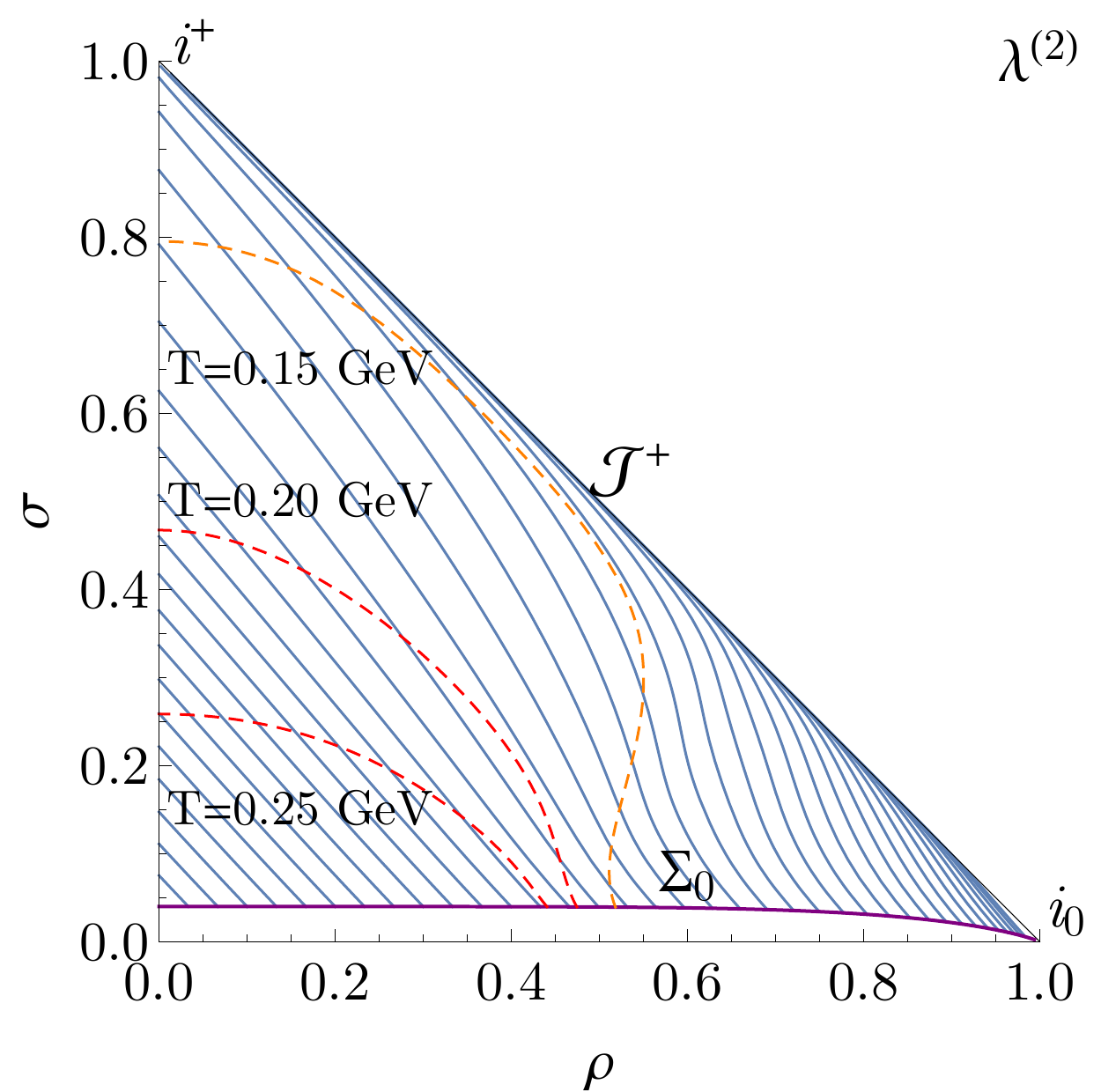}}
\subfigure[]{\label{fig:PenroseDiagram-c}
\includegraphics[width=0.40\textwidth]{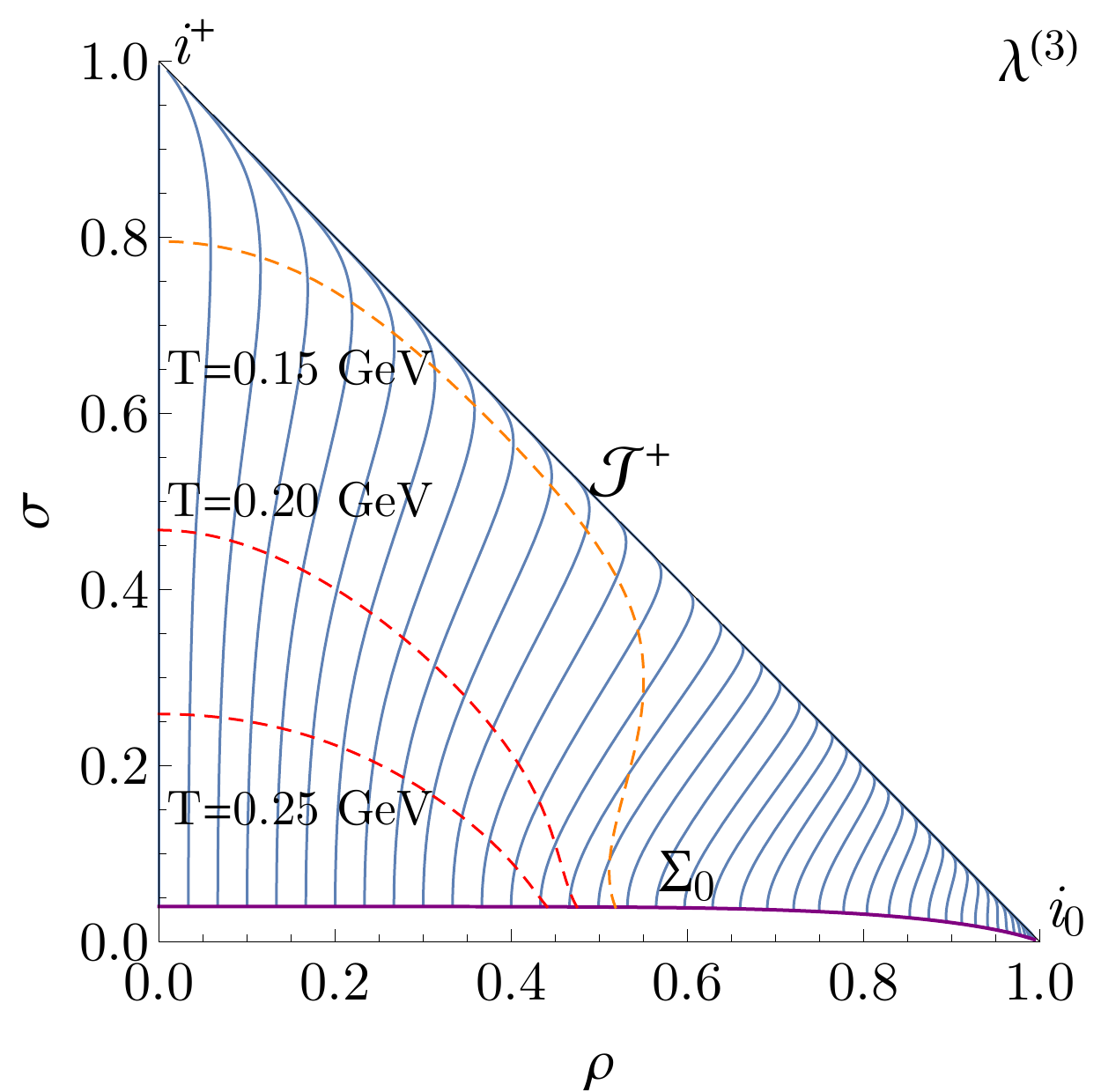}}
\caption{\label{fig:PenroseDiagram} 
Characteristic curves in the coordinates $\sigma$, $\rho$ related to Bjorken time
$\tau$ and  radius $r$ by the conformal map \eqref{eq:transformh} (see also fig. \ref{fig:coordinates_system}).
In (a) we plot characteristic curves with characteristic velocity $\lambda^{(1)}$  in \eqref{eq:eigenvaluesCharacteristics}, in (b) with characteristic velocity $\lambda^{(2)}$ and in (c) with characteristic velocity 
$\lambda^{(3)}$, which corresponds to the fluid velocity. 
For better orientation we also show curves corresponding to constant temperature as dashed lines. For a discussion of the significance of $i_0$, $i^+$ and $\mathcal{J}^+$ see figure \ref{fig:coordinates_system}. To calculate these curves 
we have used initial conditions for the shear stress according to eq.\ \eqref{eq:shear_stress_initial_condition} .  }
\end{figure}
 
 It is instructive to study in more detail the characteristic curves defined by the characteristic velocities $\lambda^{(m)}$ in  \eqref{eq:eigenvaluesCharacteristics}.
 More specific, the characteristic curves are defined as the solution of the differential equation
 \begin{equation}
 \frac{\mathrm{d} x^1}{\mathrm{d} x^0}=\lambda^{(m)}.
 \end{equation}
 For our purpose, it is particularly convenient to explore these curves in the $\sigma,\rho$ coordinate system introduced in section  \eqref{sec:CoordinateSystem}.
 We show the result in fig. \ref{fig:PenroseDiagram}. The diagram in figure 
 \ref{fig:PenroseDiagram-a} shows the characteristic curves with velocity $\lambda^{(1)}$, while figure \ref{fig:PenroseDiagram-b}  shows those 
 with characteristic velocity $\lambda^{(2)}$.  Causality demands that 
 these velocity be smaller than unity which is indeed the case.
 Figure \ref{fig:PenroseDiagram-c} shows the characteristic curves corresponding to $\lambda^{(3)}=v$. These curves can therefore also be understood as the 
 fluid flow lines. 
 For better orientation we also  show curves of constant temperature as the dashed lines in figure \ref{fig:PenroseDiagram}.
 
 \subsection{Domain of dependence and domain of influence}
 \label{sec:Domainofdependenceinfluence}
 
\begin{figure}[tbp]
\centering 
\includegraphics[width=0.4\textwidth]{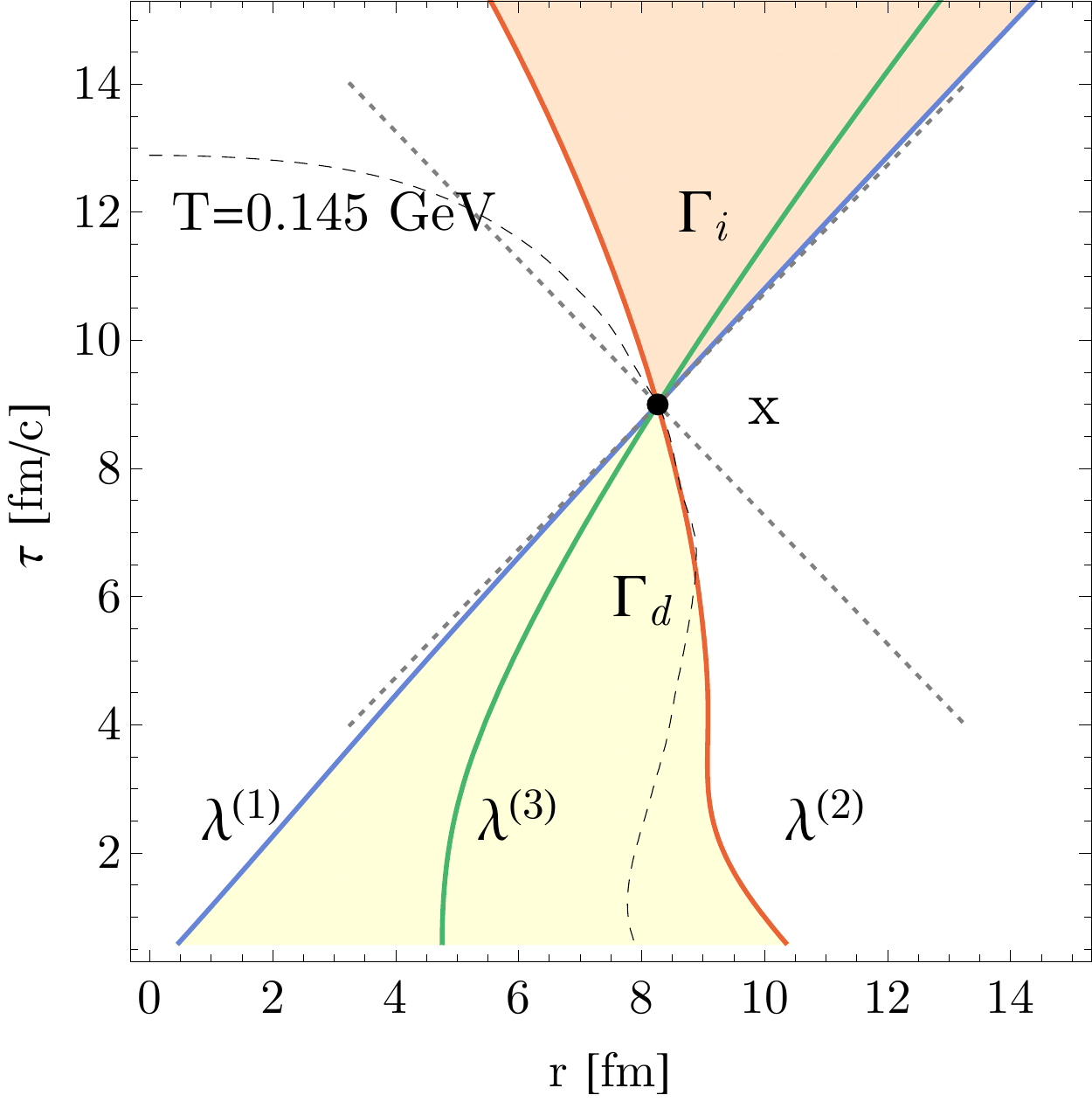}
\caption{\label{fig:lightcone} Domain of dependence $\Gamma_d$ and domain of influence $\Gamma_i$ of a space-time point $x$.  The domain of dependence $\Gamma_d$ is the region in the past of $x$ between the characteristic curves with velocities $\lambda^{(1)}$ and $\lambda^{(2)}$, respectively.  Hypothetical changes of field values in that region could modify the solution at the point $x$.
 In turn, the domain of influence $\Gamma_i$ is the region in the future of $x$
 bounded by the characteristic curves with velocities $\lambda^{(1)}$ and $\lambda^{(2)}$ where field values could be influenced by small changes at the point $x$.
 For better orientation we have also show the curve of constant temperature $T=0.145$ GeV (dashed line) and the future and the past light cones of $x$ (dotted lines).
  }
\end{figure}

 The characteristics  curves with the smallest and the largest velocity form a boundary of the region of dependence in the past of  a given space-time point $x$.
 We illustrate this in figure Fig \ref{fig:lightcone}  for a point on the line of constant temperature $T=0.145 $ GeV. The characteristic curves with velocity $\lambda^{(1)}$  and $\lambda^{(2)}$ going through this point are the boundaries
 of the region $\Gamma_d$, the region of dependence.
 This is the region inside which a (hypothetical) change of the field $\Phi$,
 in sense of  a change of initial condition, can influence the value of $\Phi$ at the point $x$.
On the other side, changing the initial condition  outside of this region has no 
influence on $\Phi(x)$, as a consequence of causality, see also appendix \ref{appC}. 
For clarity we also show the past and future light cone of the point $x$ and one can observe that indeed the domain of dependence is a subregion of the past light cone. 
 
In a similar way, one can define the domain of influence of the point $x$ as the region  in its future, bounded by the characteristic curves with smallest and largest velocity, respectively. 
As the name suggests, this is the region where field values can be influenced by a small change in the value of $\Phi(x)$ at the point $x$.
Again this region is a subregion of the corresponding light cone as required by causality.

\subsection{Causality as a bound to applicability of fluids dynamics}

\begin{figure}[tbp]
\centering 
\subfigure[]{\label{fig:Pressure Anisotropy-a}
\includegraphics[width=.45\textwidth]{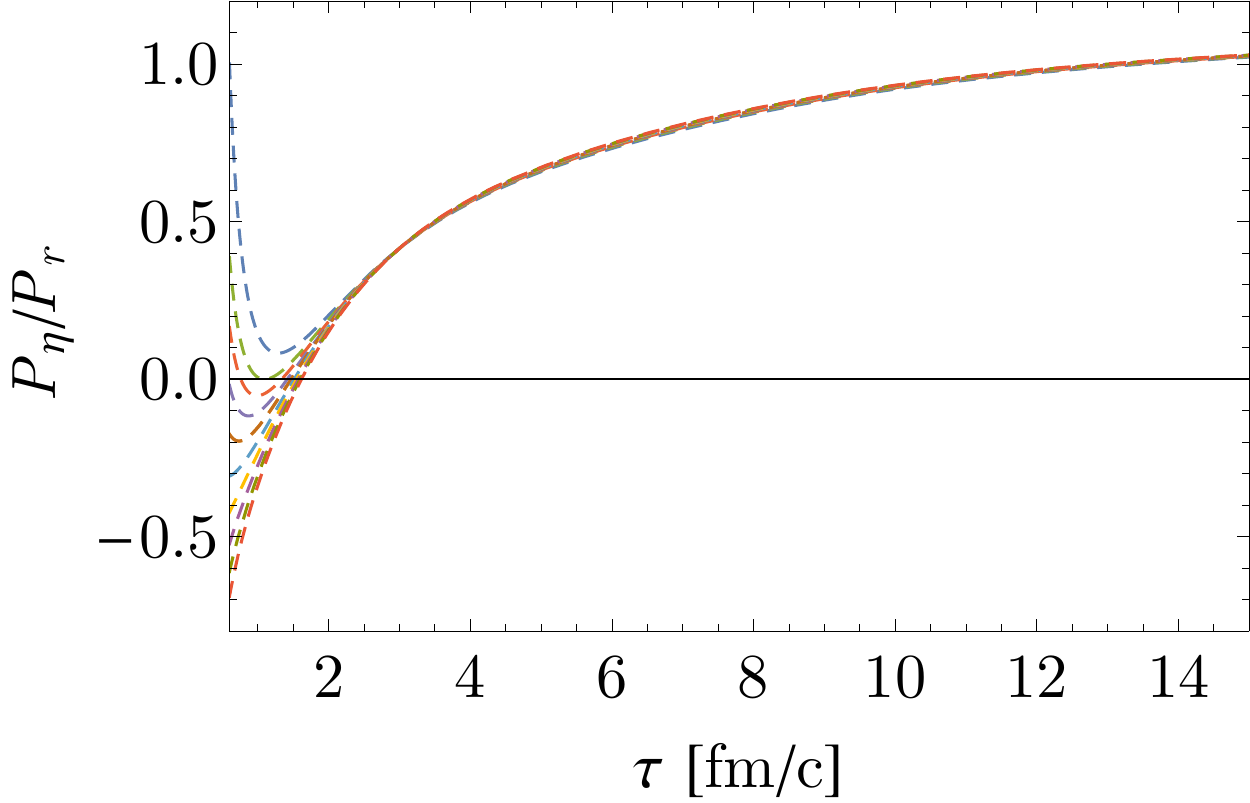}}
\subfigure[]{\label{fig:Pressure Anisotropy-b}
\includegraphics[width=.45\textwidth]{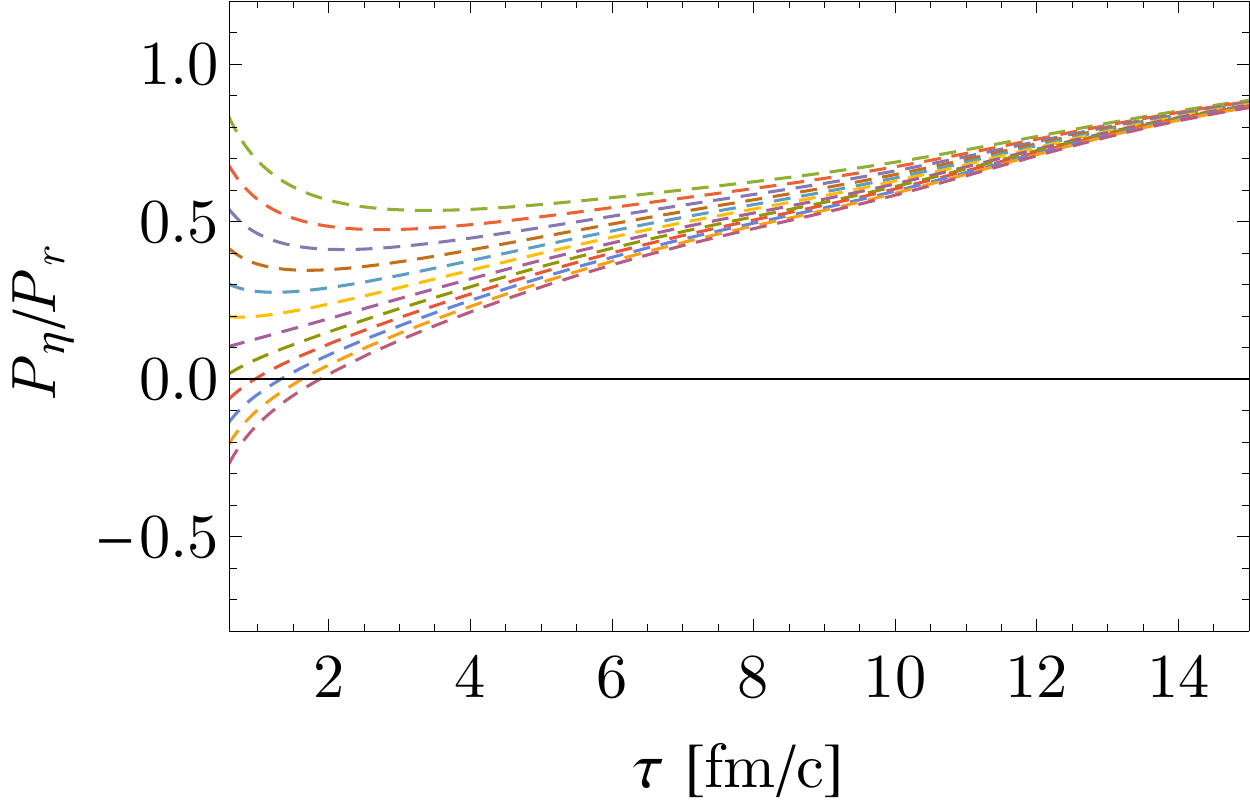}}
\caption{\label{fig:Pressure Anisotropy}
Ratio of different pressure $P_\eta/P_r$ according to \eqref{eq:PressureAniso}
in the central region of the fireball $r=0$ as a function of Bjorken time $\tau$.
We compare different initial condition for $\tilde\pi^{\eta}_\eta$ and $\tilde\pi^{\phi}_\phi$ within the range allowed by causality. For panel \ref{fig:Pressure Anisotropy-a} we have set $\tau_{\text{shear}} = 3 \eta/(\epsilon+p)$, for the panel \ref{fig:Pressure Anisotropy-b}  $\tau_{\text{shear}} = 30 \eta/(\epsilon+p)$.}
\end{figure}

As discussed in section  \ref{int} and \ref{sec2}, fluid dynamics is an effective theory for the dynamics of a quantum field theory that becomes valid for slow enough 
dynamics and small enough spatial gradients. 

Although the framework  of Israel-Stewart  theory \cite{Israel:1979wp} or the more general  DNMR  \cite{Denicol:2012cn} setup are going beyond a strict (Chapman-Enskog) derivative expansion in the thermodynamic fields and fluid velocity, it is clear that the expansion assumes small Knudsen number and inverse Reynolds number according to the definition in section \ref{sec2}. In particular, this assumes that the components of the shear stress  and bulk viscous pressure are small compared to the thermodynamic enthalpy density.

It is often stated in literature that the application of the fluid approximation needs 
an approximately isotropic energy-momentum tensor $T^{\mu\nu}$. Indeed, deviations from isotropy in the decomposition  \eqref{eq:energy_momentum_conservation} are parametrized by the shear stress.
At least superficially, a small inverse Reynolds number  $\text{Re}^{-1}$ implies also a close-to-isotropic energy-momentum tensor. However, is not clear where the bound of applicability is precisely. Oftentimes, this is discussed in terms of ratios of  
different ``pressure components", for example (assuming $\pi_{\text{bulk}}=0$ for simplicity)  
\begin{equation}
\label{eq:PressureAniso}
\frac{P_\eta}{P_r}=\frac{T^\eta_\eta}{T^{r}_r} = \frac{p+(\epsilon+p)\tilde \pi ^\eta_\eta}{p+(\epsilon+p)(\tilde \pi ^\eta_\eta+\tilde \pi^\phi_\phi)},
\end{equation}
and it is believed that such ratios should not 
deviate too much from unity in order to not leave the range of applicability of fluid 
dynamics. 
However, it is not so clear where precisely the boundary is. In particular for $ p \ll \epsilon$ the above ratio \eqref{eq:PressureAniso} could actually deviate for unity
substantially while the ratios $\tilde\pi^\eta_\eta=\pi^\eta_\eta/(\epsilon+p)$ and $\tilde\pi^\phi_\phi=\pi^\phi_\phi/(\epsilon+p)$ might be below one. 
A more severe bound for $\tilde\pi^\phi_\phi$ and $\tilde\pi^\eta_\eta$ comes from the causality constraint $|\tilde c| \le 1$  in terms of the relations \eqref{eq:modified_cs}, \eqref{eq:ideal_cs} and \eqref{eq:explicit_d}. It is interesting to investigate the range for the ratios such as the in eq.  \eqref{eq:PressureAniso} for solution of fluids dynamics that satisfy the causality bound. We have done this in fig. \ref{fig:Pressure Anisotropy-a} where we plot the ratio $P_\eta/P_r $ in the center of the fireball for different initial conditions of  
$\tilde\pi^\phi_\phi$ and $\tilde\pi^\eta_\eta$ within  the causality bound.

One observes that a rather large range of values is allowed by causality and that after a rather short time of order $1$ fm/c they approach the same attractor solution which then  evolves towards late time. The speed of approach to this attractor is actually governed by the shear stress relaxation time $\tau_{\text{shear}}$. This can be see by comparison to fig. \ref{fig:Pressure Anisotropy-b} where we have chosen  $\tau_{\text{shear}}$ to be larger by a factor 10. The approach to the attractor is accordingly slower. 
In Fig. \ref{fig:Pressure Anisotropy} we have chosen  the initialization time $\tau_0=0.6$ fm/c; for earlier initialization times $\tau_0$ the range of $P_\eta/P_r$  allowed by causality is even larger.

\section{Conclusions}
\label{concl}

We have discussed here the causality of relativistic fluid dynamics for high-energy 
nuclear collisions. 
The fluid equations we use contain all possible terms up to second order in a formal gradient expansion compatible with symmetry and with hyperbolicity  \cite{Denicol:2012cn}.
We concentrate on an expanding fireball  with longitudinal  Bjorken boost and azimuthal rotation symmetry. 
The causality of the dynamics in the reduced configuration space of Bjorken time $\tau$ and radius $r$ can be conveniently discussed in terms of characteristics.
We found five characteristic velocities for the five independent fluid fields. Three of them are degenerate and corresponds to the fluid velocity. The remaining two differ from this by a modified sound velocity (see eq. \eqref{eq:modified_cs}, \eqref{eq:ideal_cs} and \eqref{eq:explicit_d}) which contains the standard ideal fluid velocity of sound but also nonlinear modifications due to the dissipative terms. In particular, the modified sound velocity depends not only on thermodynamic variables and transport properties, but also on the components of the shear stress tensor and bulk viscous pressure. Causality is in this sense state-dependent.

In a subsequent  step  we investigate whether the modified sound velocity remains below the speed of light for characteristic situations that occur in the context of high-energy nuclear collisions. 
Depending on the initial condition, we found that the causality constraint can be violated at early times, which indicates a violation of the  applicability conditions for the fluid approximation.
In particular, such violations occur for so-called Navier-Stokes initial conditions that have often been used in phenomenological investigations. We also show that modifications of these initial conditions lead to a dynamical evolution that does not violate the causality constraint. Finally, we investigate the ratio of longitudinal 
and radial pressure in the center of the fireball for a range of initial conditions that does not lead to any violations of causality. We find that this ratio can vary substantially at early times but approaches an attractor solution quickly  at late times where the  ``speed of approach" is determined  by the shear stress relaxation
time as expected.

The most important conclusion from our findings is that causality poses a bound on the applicability of relativistic fluid dynamics. 
This helps for concrete phenomenological applications of the formalism but also  
improves the more conceptual understanding of its foundations.

\section*{Acknowledgements}
This work is part of and supported by the DFG Collaborative Research Centre ``SFB 1225 (ISOQUANT)''.

\appendix
\section{Generalized Bjorken coordinates}
\label{appA}

We collect here some useful relations for the generalized Bjorken coordinates introduced in section \ref{sec:CoordinateSystem}. Let us recall the Minkowski space metric
\begin{equation}
ds^2=h'(\rho +\sigma)h'(\rho -\sigma)\left[-d\sigma^2+d\rho^2\right] + \tfrac{1}{4}[h(\rho +\sigma)+h(\rho -\sigma)]^2 d \phi^2 + \tfrac{1}{4}[h(\rho +\sigma)-h(\rho -\sigma)]^2 d\eta^2.
\label{eq:appA2}
\end{equation}
The coordinates $\sigma$ and $\rho$ are related to the more familiar longitudinal proper time $\tau=\sqrt{t^2-z^2}$ and transverse radius $r=\sqrt{x^2+y^2}$ by
\begin{equation}
\begin{split}
r=\frac{1}{2}\left[h(\rho +\sigma)+h(\rho -\sigma)\right], & \quad\quad \tau=\frac{1}{2}\left[h(\rho +\sigma)-h(\rho -\sigma)\right],\\
\rho=\frac{1}{2}\left[h^{-1}(r +\tau)+h^{-1}(r -\tau)\right], & \quad\quad \sigma=\frac{1}{2}\left[h^{-1}(r +\tau)-h^{-1}(r -\tau)\right].
\end{split}
\end{equation}

The non-vanishing Christoffel symbols are
\begin{equation}
\begin{split}
\Gamma^0_{00} & = \Gamma^0_{11}= \Gamma^1_{01}=  \Gamma^1_{10} = \frac{1}{2} \left( \frac{h^{\prime\prime}(\rho+\sigma)}{h^\prime(\rho+\sigma)} - \frac{h^{\prime\prime}(\rho-\sigma)}{h^\prime(\rho-\sigma)} \right) = \partial_0 \, \text{ln}\left( \sqrt{g_{11}} \right), \\
\Gamma^1_{11} & = \Gamma^1_{00}= \Gamma^0_{01}=  \Gamma^0_{10} = \frac{1}{2} \left( \frac{h^{\prime\prime}(\rho+\sigma)}{h^\prime(\rho+\sigma)} + \frac{h^{\prime\prime}(\rho-\sigma)}{h^\prime(\rho-\sigma)} \right) = \partial_1 \, \text{ln}\left( \sqrt{g_{11}} \right), \\
\Gamma^0_{22} & = \frac{\left(h(\rho+\sigma) + h(\rho-\sigma)  \right) \left( h^\prime(\rho+\sigma) - h^\prime(\rho-\sigma) \right)}{4 h^\prime(\rho+\sigma) h^\prime(\rho-\sigma)}, \\
\Gamma^0_{33} & = \frac{\left(h(\rho+\sigma) - h(\rho-\sigma)  \right) \left( h^\prime(\rho+\sigma) + h^\prime(\rho-\sigma) \right)}{4 h^\prime(\rho+\sigma) h^\prime(\rho-\sigma)}, \\
\Gamma^1_{22} & = - \frac{\left(h(\rho+\sigma) + h(\rho-\sigma)  \right) \left( h^\prime(\rho+\sigma) + h^\prime(\rho-\sigma) \right)}{4 h^\prime(\rho+\sigma) h^\prime(\rho-\sigma)}, \\
\Gamma^1_{33} & = - \frac{\left(h(\rho+\sigma) - h(\rho-\sigma)  \right) \left( h^\prime(\rho+\sigma) - h^\prime(\rho-\sigma) \right)}{4 h^\prime(\rho+\sigma) h^\prime(\rho-\sigma)}, \\
\Gamma^2_{02} & = \Gamma^2_{20} = \frac{h^\prime(\rho+\sigma) - h^\prime(\rho-\sigma)}{h(\rho+\sigma)+h(\rho-\sigma)} = \partial_0 \, \text{ln}\left( \sqrt{g_{22}} \right),\\
\Gamma^2_{12} & = \Gamma^2_{21} = \frac{h^\prime(\rho+\sigma) + h^\prime(\rho-\sigma)}{h(\rho+\sigma)+h(\rho-\sigma)}= \partial_1 \, \text{ln}\left( \sqrt{g_{22}} \right), \\
\Gamma^3_{03} & = \Gamma^3_{30} = \frac{h^\prime(\rho+\sigma) + h^\prime(\rho-\sigma)}{h(\rho+\sigma)-h(\rho-\sigma)}= \partial_0 \, \text{ln}\left( \sqrt{g_{33}} \right), \\
\Gamma^3_{13} & = \Gamma^3_{31} = \frac{h^\prime(\rho+\sigma) - h^\prime(\rho-\sigma)}{h(\rho+\sigma) - h(\rho-\sigma)}= \partial_1 \, \text{ln}\left( \sqrt{g_{33}} \right) ,
\end{split}
\end{equation}
and the square root of the metric determinant is
\begin{equation}
\sqrt{g} = \sqrt{- \text{det} \, g_{\mu\nu}} =\frac{1}{4} h^\prime(\rho+\sigma) h^\prime(\rho-\sigma) \left[ h(\rho+\sigma) + h(\rho-\sigma) \right] \left[ h(\rho+\sigma) - h(\rho-\sigma) \right].
\end{equation}
 Note that in terms of $\sigma$ and $\rho$, the half plane $0\leq \tau, r < \infty$ corresponds to $0 \leq \sigma, \rho < 1$ with $\rho + \sigma < 1$. However, it is sometimes useful to analytically continue to negative $\rho $ with $|\rho | < 1-\sigma$. One can obviously recover the original Bjorken coordinate system with the trivial choice $h(x)=x$ such that $\sigma=\tau$ and $\rho=r$. 

In a 1+1 dimensional situation it is convenient to parametrize the fluid velocity as
\begin{equation}
u^\mu = \left( \frac{\text{cosh}(\chi)}{\sqrt{g_{11}}}, \frac{\text{sinh}(\chi)}{\sqrt{g_{11}}}, 0, 0 \right),
\label{eq:fluidvelocityparametrization}
\end{equation}
with radial fluid rapidity field $\chi$. The projector orthogonal to the fluid velocity is
\begin{equation}
\Delta^\mu_{\;\;\nu} = \delta^\mu_{\;\;\nu} + u^\mu u_\nu  = \begin{pmatrix} -\text{sinh}^2(\chi) & \text{cosh}(\chi) \text{sinh}(\chi) & 0 & 0 \\ -  \text{cosh}(\chi) \text{sinh}(\chi) & \text{cosh}^2(\chi) & 0 & 0 \\ 0 & 0 & 1 & 0 \\ 0 & 0 & 0 & 1 \end{pmatrix}.
\label{eq:parametrizationFluidVelocity}
\end{equation}
The fluid velocity divergence is 
\begin{equation}
\sqrt{g_{11}} \; \nabla_\mu u^\mu = \partial_0 \, \text{cosh}(\chi) + \partial_1 \, \text{sinh}(\chi) + \text{cosh}(\chi) \, \partial_0 \, \text{ln}\left( \sqrt{g_{11}}\sqrt{g_{22}} \sqrt{g_{33}} \right) + \text{sinh}(\chi) \, \partial_1 \, \text{ln}\left( \sqrt{g_{11}}\sqrt{g_{22}} \sqrt{g_{33}} \right).
\end{equation}
We also need the tensor $\nabla_\mu u^\nu$. It has the following non-vanishing components
\begin{equation}
\begin{split}
\sqrt{g_{11}}\, \nabla_0 u^0 = & \partial_0 \, \text{cosh}(\chi) + \text{sinh}(\chi) \, \partial_1 \, \text{ln}\left( \sqrt{g_{11}} \right), \\
\sqrt{g_{11}}\, \nabla_0 u^1 = & \partial_0 \, \text{sinh}(\chi) + \text{cosh}(\chi) \, \partial_1 \, \text{ln}\left( \sqrt{g_{11}} \right), \\
\sqrt{g_{11}}\, \nabla_1 u^0 = & \partial_1 \, \text{cosh}(\chi) + \text{sinh}(\chi) \, \partial_0 \, \text{ln}\left( \sqrt{g_{11}} \right), \\
\sqrt{g_{11}}\, \nabla_1 u^1 = & \partial_1 \, \text{sinh}(\chi) + \text{cosh}(\chi) \, \partial_0 \, \text{ln}\left( \sqrt{g_{11}} \right), \\
\sqrt{g_{11}}\, \nabla_2 u^2 = & \text{cosh}(\chi) \, \partial_0 \, \text{ln}\left( \sqrt{g_{22}} \right) + \text{sinh}(\chi) \, \partial_1 \, \text{ln}\left( \sqrt{g_{22}} \right), \\
\sqrt{g_{11}}\, \nabla_3 u^3 = & \text{cosh}(\chi) \, \partial_0 \, \text{ln}\left( \sqrt{g_{33}} \right) + \text{sinh}(\chi) \, \partial_1 \, \text{ln}\left( \sqrt{g_{33}} \right).
\end{split}
\end{equation}
The shear stress $\pi^{\mu\nu}$ has only two independent components and for convenience they can be chosen as the dimensionless ratios $\tilde \pi^{\eta}_{\eta}=\pi^3_{\;\;3}/(\epsilon+p)$ and $\tilde \pi^\phi_{\phi}=\pi^2_{\;\;2}/(\epsilon+p)$. The other non-vanishing components are related to these by
\begin{equation}
\begin{split}
& \pi^0_{\;\;0}=\text{sinh}^2(\chi) \left( \tilde \pi^\phi_\phi + \tilde \pi^\eta_\eta \right)(\epsilon+p), \quad\quad\quad \pi^1_{\;\;1} = - \text{cosh}^2(\chi) \left( \tilde \pi^\phi_\phi + \tilde \pi^\eta_\eta \right)(\epsilon+p), \\
& \pi^1_{\;\;0} = - \pi^0_{\;\;1} = \text{cosh}(\chi) \, \text{sinh}(\chi) \left( \tilde \pi^\phi_\phi + \tilde \pi^\eta_\eta \right)(\epsilon+p).
\end{split}
\label{eq:parametrizationShearStress}
\end{equation}
We also parametrize the bulk viscous pressure by a dimensionless variable as $\pi_\text{bulk} = \tilde \pi_\text{bulk}(\epsilon+p)$.

\section{Equations of motion}
\label{app:EOM}

In this appendix we compile the relativistic fluid dynamic equations of motion for a central heavy ion collision with longitudinal and transverse expansion. The equations are partial differential equations involving a time coordinate $x^0$ and a radial coordinate $x^1$. The evolution equations for energy density and fluid velocity follow directly from energy-momentum conservation. The energy density, pressure and other thermodynamic and transport properties close to local equilibrium can be expressed in terms of a single independent thermodynamic variable, which we take to be temperature. 

Using standard thermodynamic relations such as $\partial p/\partial T= s$, $\partial \epsilon/\partial T= s/c_s^2$  and dividing by the enthalpy $w =\epsilon+p$,  one finds from the energy equation in \eqref{eq:energy_momentum_conservation},
\begin{equation}
\begin{split}
 & \frac{1}{c_s^2} \left[ \text{cosh}(\chi) \partial_0 \text{ln} \,T+ \text{sinh}(\chi) \partial_1 \text{ln}\, T \right] \\
 & + \left(1+\tilde{\pi}_\text{bulk} \right) \left[  \text{sinh}(\chi)\, \partial_0\chi + \text{cosh}(\chi) \, \partial_1\chi + \text{cosh}(\chi) \, \partial_0 \, \text{ln}\left( \sqrt{g_{11}} \sqrt{g_{22}} \sqrt{g_{33}} \right) + \text{sinh}(\chi) \, \partial_1 \, \text{ln}\left( \sqrt{g_{11}} \sqrt{g_{22}} \sqrt{g_{33}} \right) \right] \\
 & - \tilde{\pi}^\phi_\phi \left[ \text{sinh}(\chi) \, \partial_0 \chi + \text{cosh}(\chi) \, \partial_1 \chi  + \text{cosh}(\chi) \, \partial_0 \, \text{ln}\left( \sqrt{g_{11}} / \sqrt{g_{22}} \right)
 + \text{sinh}(\chi) \, \partial_1 \, \text{ln}\left( \sqrt{g_{11}} / \sqrt{g_{22}} \right)  \right] \\
  & - \tilde{\pi}^\eta_\eta \left[ \text{sinh}(\chi) \,  \partial_0 \chi +  \text{cosh}(\chi) \, \partial_1 \chi + \text{cosh}(\chi) \, \partial_0 \, \text{ln}\left( \sqrt{g_{11}} / \sqrt{g_{33}} \right)
 + \text{sinh}(\chi) \, \partial_1 \, \text{ln}\left( \sqrt{g_{11}} / \sqrt{g_{33}} \right)  \right] = 0.
\end{split}
\end{equation}
The equation for the radial fluid velocity yields in terms of the parametrization \eqref{eq:parametrizationFluidVelocity},
\begin{equation}
\begin{split}
& \left(1 + \tilde{\pi}_\text{bulk} \right) \left[ \text{cosh}(\chi) \, \partial_0 \chi + \text{sinh}(\chi) \, \partial_1\chi  + \text{sinh}(\chi) \, \partial_0 \, \text{ln}\left( \sqrt{g_{11}} \right) + \text{cosh}(\chi) \, \partial_1 \, \text{ln}\left( \sqrt{g_{11}} \right) \right] \\
& +\left[1+\tilde\pi_\text{bulk} -\tilde\pi_\phi^\phi- \tilde\pi^\eta_\eta +\frac{1}{c_s^2}(\tilde\pi_\text{bulk} -\tilde\pi_\phi^\phi- \tilde\pi^\eta_\eta) \right] \left[ \text{sinh}(\chi) \, \partial_0 \text{ln}\, T + \text{cosh}(\chi) \, \partial_1 \text{ln}\, T \right] 
\\
&+ \left[ \text{sinh}(\chi) \, \partial_0 \, \tilde{\pi}_\text{bulk} + \text{cosh}(\chi) \, \partial_1 \, \tilde{\pi}_\text{bulk} 
 -   \text{sinh}(\chi) \, \partial_0 \, \tilde{\pi}^\phi_\phi -\text{cosh}(\chi) \, \partial_1 \, \tilde{\pi}^\phi_\phi - \text{sinh}(\chi) \, \partial_0 \, \tilde{\pi}^\eta_\eta - \text{cosh}(\chi) \, \partial_1 \, \tilde{\pi}^\eta_\eta \right] \\
& - \tilde{\pi}^\phi_\phi \left[ \text{cosh}(\chi) \, \partial_0 \chi+ \text{sinh}(\chi) \, \partial_1\chi  + \text{sinh}(\chi) \, \partial_0 \, \text{ln}\left( \sqrt{g_{11}} g_{22} \sqrt{g_{33}} \right)
+ \text{cosh}(\chi) \, \partial_1 \, \text{ln}\left( \sqrt{g_{11}} g_{22} \sqrt{g_{33}} \right) \right] \\
& - \tilde{\pi}^\eta_\eta \left[ \text{cosh}(\chi) \, \partial_0 \chi + \text{sinh}(\chi) \, \partial_1 \chi + \text{sinh}(\chi) \, \partial_0 \, \text{ln}\left( \sqrt{g_{11}} \sqrt{g_{22}} g_{33} \right)
+ \text{cosh}(\chi) \, \partial_1 \, \text{ln}\left( \sqrt{g_{11}} \sqrt{g_{22}} g_{33} \right) \right] = 0.
\end{split}
\end{equation}
These equations get supplemented by the evolution equations for the shear stress. From \eqref{eq:shear_tensor_conservation_DNMR} and the parametrization \eqref{eq:parametrizationShearStress} one finds for $\tilde \pi^\phi_\phi$ the equation
\begin{equation}
\begin{split}
& \tau_\text{shear} \left[ \text{cosh}(\chi) \, \partial_0 \, \tilde\pi^\phi_\phi + \text{sinh}(\chi) \, \partial_1 \, \tilde\pi^\phi_\phi \right] +\tau_\text{shear} \tilde\pi^\phi_\phi \left(1+\frac{1}{c_s^2}\right)
\left[ \text{cosh}(\chi) \, \partial_0 \,\text{ln}\, T + \text{sinh}(\chi) \, \partial_1 \, \text{ln}\, T \right] 
\\
&+ \sqrt{g_{11}} \, \tilde\pi^\phi_\phi\left(1-\frac{ \varphi_6}{w}  \tilde\pi_{\text{bulk}}  -\frac{\varphi_7 }{w} \tilde\pi^\phi_\phi \right)
 -\left(\frac{2\eta}{w}+   \tau_{\pi\pi}  \tilde\pi^\phi_\phi- \lambda_{\pi\Pi} \tilde\pi_{\text{bulk}}\right) \\&
\times \frac{1}{3}\left[ \text{sinh}(\chi) \, \partial_0 \chi + \text{cosh}(\chi) \partial_1 \chi + \text{cosh}(\chi) \, \partial_0 \, \text{ln}\left( \sqrt{g_{11}} \sqrt{g_{33}} / g_{22} \right) 
+ \text{sinh}(\chi) \, \partial_1 \, \text{ln}\left( \sqrt{g_{11}} \sqrt{g_{33}} / g_{22} \right) \right] \\
& + \tilde\pi^\phi_\phi  \delta_{\pi\pi} \left[ \text{sinh}(\chi) \, \partial_0 \chi +  \text{cosh}(\chi) \, \partial_1 \chi + \text{cosh}(\chi) \, \partial_0 \, \text{ln}\left( \sqrt{g_{11}}\sqrt{g_{22}} \sqrt{g_{33}} \right) + \text{sinh}(\chi) \, \partial_1 \, \text{ln}\left( \sqrt{g_{11}}\sqrt{g_{22}} \sqrt{g_{33}} \right) \right]
= 0.
\end{split}
\end{equation}

%
In a similar way, for $\tilde \pi^\eta_\eta$ one obtains,
\begin{equation}
\begin{split}
& \tau_\text{shear} \left[ \text{cosh}(\chi) \, \partial_0 \, \tilde\pi^\eta_\eta + \text{sinh}(\chi) \, \partial_1 \, \tilde\pi^\eta_\eta \right] 
+ \tau_\text{shear} \left(1+\frac{1}{c_s^2}\right)
 \tilde\pi^\eta_\eta \left[ \text{cosh}(\chi) \, \partial_0 \, \text{ln} \, T + \text{sinh}(\chi) \, \partial_1 \, \text{ln} \, T \right] 
\\&+ \sqrt{g_{11}} \, \tilde\pi^\eta_\eta\left(1-\frac{  \varphi_6}{w}\tilde\pi_{\text{bulk}}-\frac{  \varphi_7 }{w} \tilde\pi_{\eta}^\eta\right)-\left(\frac{2\eta}{w} +   \tau_{\pi\pi} \tilde\pi^{\eta}_\eta-  \lambda_{\pi\Pi} \tilde\pi_{\text{bulk}} \right) \\
& \times \frac{1}{3} \left[ \text{sinh}(\chi) \, \partial_0 \chi  + \text{cosh}(\chi) \, \partial_1 \chi  + \text{cosh}(\chi) \, \partial_0 \, \text{ln}\left( \sqrt{g_{11}} \sqrt{g_{22}} / g_{33} \right) 
+ \text{sinh}(\chi) \, \partial_1 \, \text{ln}\left( \sqrt{g_{11}} \sqrt{g_{22}} / g_{33} \right) \right] \\
& + \tilde\pi^\eta_\eta  \delta_{\pi\pi}  \left[ \text{sinh}(\chi) \, \partial_0 \chi +  \text{cosh}(\chi) \, \partial_1 \chi + \text{cosh}(\chi) \, \partial_0 \, \text{ln}\left( \sqrt{g_{11}}\sqrt{g_{22}} \sqrt{g_{33}} \right) + \text{sinh}(\chi) \, \partial_1 \, \text{ln}\left( \sqrt{g_{11}}\sqrt{g_{22}} \sqrt{g_{33}} \right) \right]
= 0.
\end{split}
\end{equation}
Finally, the evolution equation for the bulk viscous pressure follows from \eqref{eq:bulk_pressure_conservation_DNMR} as
\begin{equation}
\begin{split}
& \tau_\text{bulk} \left[ \text{cosh}(\chi) \, \partial_0 \, \tilde\pi_\text{bulk} + \text{sinh}(\chi) \, \partial_1 \, \tilde\pi_\text{bulk}\right] 
+ \tau_\text{bulk}  \left(1+\frac{1}{c_s^2}\right)\tilde\pi_\text{bulk} \left[ \text{cosh}(\chi) \, \partial_0 \, \text{ln} \, T + \text{sinh}(\chi) \, \partial_1 \, \text{ln} \, T\right] 
\\&+ \sqrt{g_{11}} \; \left[\tilde\pi_\text{bulk} -\frac{  \varphi_1 }{w} \tilde\pi_\text{bulk}^2 -\frac{2  \varphi_3 }{w} ((\tilde\pi_{\phi}^\phi )^2+(\tilde\pi_{\eta}^\eta)^2+\tilde\pi_{\phi}^\phi \tilde\pi^{\eta}_\eta)\right] \\
& +\left( \frac{\zeta}{w}+ \delta_{\Pi\Pi}  \tilde\pi_\text{bulk} \right)\left[ \text{sinh}(\chi) \, \partial_0 \chi + \text{cosh}(\chi) \, \partial_1 \chi + \text{cosh}(\chi) \, \partial_0 \, \text{ln}\left( \sqrt{g_{11}}\sqrt{g_{22}} \sqrt{g_{33}} \right) + \text{sinh}(\chi) \, \partial_1 \, \text{ln}\left( \sqrt{g_{11}}\sqrt{g_{22}} \sqrt{g_{33}} \right) \right] \\
&+  \lambda_{\Pi\pi} \tilde\pi^\phi_\phi
\left[ \text{sinh}(\chi) \, \partial_0 \chi + \text{cosh}(\chi) \, \partial_1 \chi + \text{cosh}(\chi) \, \partial_0 \, \text{ln}\left( \sqrt{g_{11}}/\sqrt{g_{22}}  \right) + \text{sinh}(\chi) \, \partial_1 \, \text{ln}\left( \sqrt{g_{11}}/\sqrt{g_{22}}\right) \right]
\\
&+ \lambda_{\Pi\pi}  \tilde\pi^\eta_\eta
\left[ \text{sinh}(\chi) \, \partial_0 \chi + \text{cosh}(\chi) \, \partial_1 \chi + \text{cosh}(\chi) \, \partial_0 \, \text{ln}\left( \sqrt{g_{11}}/\sqrt{g_{33}}  \right) + \text{sinh}(\chi) \, \partial_1 \, \text{ln}\left( \sqrt{g_{11}}/\sqrt{g_{33}}\right) \right]
= 0.
\end{split}
\end{equation}

\section{Energy inequalities and uniqueness of the solution }
\label{appC}

\begin{figure}[tbp]
\centering 
\includegraphics[width=0.4\textwidth]{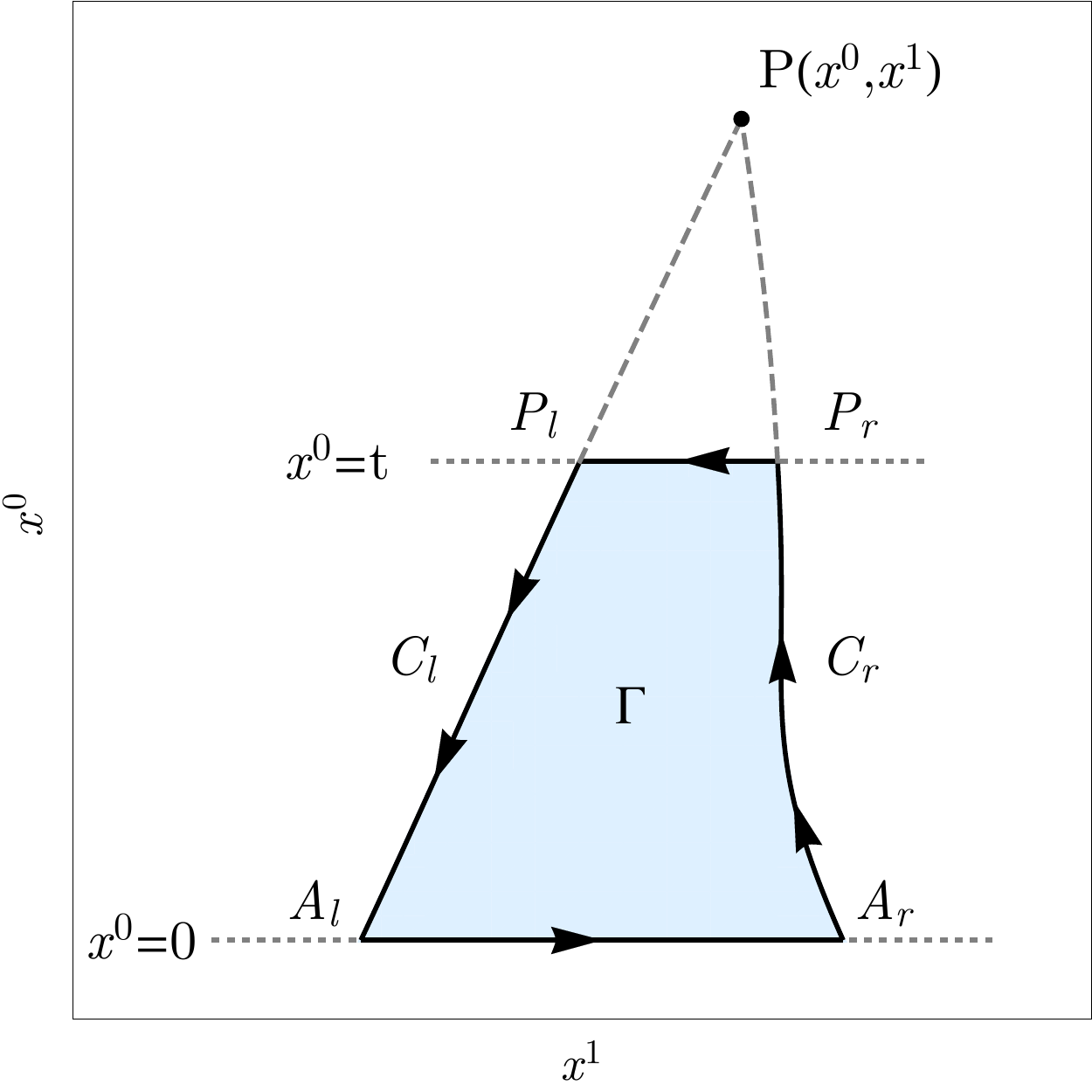}
\caption{\label{fig:CausalPlot}
Domain of dependence of the point $P$. Here,
$C_l$ is the characteristic curve with largest velocity and $C_r$ with the smallest velocity meeting at $P$, the line $x^0 = 0$ is the where the initial data are provided and $x^0 = t$ is an intermediate fixed time line.}
\end{figure}

In this appendix we explain in more details the notion of causality of hyperbolic equations and its relation with characteristic curves. As we discuss in the main text, the evolution is causal if the domain of dependence $\Gamma_d$ of a given point $P=(x^0,x^1)$ is contained in the past light cone of that point. 
The domain of dependence $\Gamma_d$ is defined as the region in the past  of the point $P$ bounded by the two extreme characteristic curves, i.e. the two integral curves with maximal and minimal velocities. 
Any perturbation of the initial data  that vanishes in this region does not 
change the value of the solution at the point $P$.
This property is due to the existence of a so-called ``energy inequality" that bounds 
the spatial average of the solution to the spatial average of the initial data. 
To be more specific consider a perturbation $\delta \Phi$, solution of the linearized version of the partial differential equation  \eqref{eq:quasilinearmatix} around a generic background $\Phi_0$, and consider the domain of dependence of the point $P$ shown in figure  \ref{fig:CausalPlot}: $C_l$ is the characteristic curves with largest velocity and $C_r$ with the smallest velocity, the line $x^0=0$ is the where the initial data are provided and $x^0=t$ is an intermediate fixed time line. 
With the perturbation $\delta \Phi$ it is possible to construct the new variables 
$\delta J^{(m)}=w^{(m)}_i\delta \Phi_i$ where $w^{(m)}_i$ are the left eigenvectors evaluated on the background solution \eqref{eq:left_eigenvectors}. 
With these variables it is possible to formulate the energy inequality (see below for a proof)
\begin{equation}
0\le   \frac{1}{2}\int_{P_l}^{P_r}\mathrm{d}x^1 \delta J^m\delta J^m
 \le 
 e^{2\mu t}  \frac{1}{2}\int_{A_l}^{A_r}\mathrm{d}x^1 \delta J^m\delta J^m,
\end{equation}
where $A_l$, $A_r$, $P_l$ and $P_r$ are the  intersection points of the characteristic curves with constant time lines shown  int the figure \ref{fig:CausalPlot}, and $\mu $ is some positive constant.
The first consequence of this inequality is that the solution of the linearized equation $\delta \Phi$ can grow at most exponential with respect to the initial data. 
The second consequence, and most interesting for our purpose, is that if the initial data vanish, i.e.
\begin{equation}
\frac{1}{2}\int_{A_l}^{A_r}\mathrm{d}x^1 \delta J^m\delta J^m=0,
\end{equation}
then 
\begin{equation}
0\le 
\frac{1}{2}\int_{P_l}^{P_r}\mathrm{d}x^1 \delta J^m\delta J^m \le 0,
\end{equation}
for all times $x^0=t$, and in particular at the point $P$, therefore  $\delta \Phi(P)=0$.
As a consequence, any perturbation $\delta\phi$ outside of the domain of dependence can not 
modify the value of the solution at the point $P$, to linear order.

 In the rest of this appendix we  recall briefly the proof of the energy inequalities that can be found in 
\cite{1953mmp..book.....C}  applied to equation \eqref{eq:quasilinearmatix}.
Consider a small perturbation $\delta \Phi$ around a generic solution $\Phi^0(x) $ of \eqref{eq:quasilinearmatix},
\begin{equation}
\label{eq:perturbation}
 \partial_0 \delta \Phi_j + M_{jk} \partial_1 \delta\Phi_k + S_{jk}\delta\Phi_k = 0.
\end{equation}
where 
\begin{equation}
\begin{split}
M_{jk}(x)= (A^{-1}B)_{jk}\Big|_{\Phi=\Phi^0(x)},\\
S_{jk}(x)=\frac{\partial  (A^{-1}B)_{ji} }{\partial \Phi_k}\partial_1 \delta\Phi^0_i+ \frac{\partial  (A^{-1}C)_{j} }{\partial \Phi_k}\Big|_{\Phi=\Phi^0(x)}.
\end{split}
\end{equation}
This set of equations is linear and hyperbolic and the matrix $M$ has a 
set of left eigenvector given by \eqref{eq:left_eigenvectors}. Eq.\ \eqref{eq:perturbation} is not symmetric-hyperbolic because $M$ in general is not a symmetric matrix. However, we can introduce the variables $\delta J^{(m)}$, that diagonalized the derivative part of the equations if we multiply  \eqref{eq:perturbation} with $w_j^{(m)}(x)$ from the left,
\begin{equation}
\begin{split}
\delta J^{(m)} =w_j^{(m)}(x) \delta \Phi_j,
\\
 \partial_0 \delta J^{(m)}+ \lambda^{(m)} \partial_1 \delta J^{(m)} + w_j^{(m)} S_{jk}\delta\Phi_k +( \partial_0 w_j^{(m)} +\lambda^{(m)} \partial_1
 w_j^{(m)} 
 ) \delta \Phi_j= 0.
 \end{split}
\end{equation}
The source term can be expressed in terms of $\delta J^{(m)}$ as well using the right eigenvectors $r_j^{(m)}$ of $M_{jk}$ normalized such that
$$
\sum_{j}w_j^{(m)}r_j^{(n)}=\delta^{mn}.
$$
This yields
\begin{equation}
\begin{split}
 \partial_0 \delta J^{(m)}+ \lambda^{(m)} \partial_1 \delta J^{(m)} + F_{mn} \delta J^{(n)}= 0,
\end{split}
\end{equation}
with $F_{mn}$ defined by 
$$
F_{mn}(x)=w_j^{(m)} S_{jk}r_k^{(n)} + \partial_0 w_j^{(m)} r_j^{(n)} +\lambda^{(m)} \partial_1
 w_j^{(m)} r_j^{(n)} . 
$$

The linearized equations are now written  in a form where the matrix coupled to the radial derivative is diagonal and it is possible to write
\begin{equation}
\frac{1}{2}\partial_0(\delta J,\delta J)+\frac{1}{2}\partial_1(\delta J,\Lambda \delta J)+(\delta J, [F-\frac{1}{2}\partial_1 \Lambda ] \delta J )=0,
\end{equation}
where $\Lambda=\text{diag}(\lambda^{(1)},\lambda^{(2)},\cdots,\lambda^{(n)})$ and the scalar product $(\cdot ,\cdot)$ is the Euclidean one, defined on the vector space of the perturbations.
The unknown vector $\delta J$ can be rescaled, $\delta J =e^{\mu x^0} \overline{\delta J}$ which leads to
\begin{equation}
\frac{1}{2}\partial_0(\overline{\delta J} ,\overline{\delta J} )+\frac{1}{2}\partial_1(\overline{\delta J} ,\Lambda \overline{\delta J} )=(\overline{\delta J} , [-\mu I - F+\frac{1}{2}\partial_1 \Lambda ] \overline{\delta J}  ).
\end{equation}
The quadratic form on the right hand side can be taken as a negative definite, if 
we select a sufficiently large value for the constant $\mu$. Consequently we can write an inequities for the left hand side,
\begin{equation}
\label{eq:inequality}
\frac{1}{2}\partial_0(\overline{\delta J} ,\overline{\delta J} )+\frac{1}{2}\partial_1(\overline{\delta J} ,\Lambda \overline{\delta J} )\le 0.
\end{equation}

For a given point $P=(x^0,x^1)$ is possible to compute the characteristic curves
$x^1_m(x^0)$ solving the differential equation $\mathrm{d} x^1/\mathrm{d}x^0=\lambda^{(m)}$, this is possible because the eigenvalue $\lambda^{(m)}$ depends only on the background solution. Consider now the trapezoidal domain $\Gamma$ bound by the lines $x^0=0$, $x^0=t$ and the two outer characteristic curves that arrive at the point $P$ (see figure \ref{fig:CausalPlot}).
Integrating the inequality \eqref{eq:inequality} over this domain, we obtain the inequality
\begin{equation}
\label{eq:integral_inequality}
\int_{\Gamma}\mathrm{d}^2 x \left\{ \frac{1}{2}\partial_0(\overline{\delta J} ,\overline{\delta J} )+\frac{1}{2}\partial_1(\overline{\delta J} ,\Lambda \overline{\delta J} ) \right\} \le 0.
\end{equation}

Using Gauss's theorem, one now rewrites this as a line integral over the boundary $\partial\Gamma$ (see fig.\ \ref{fig:CausalPlot}), leading to
\begin{equation}
\label{eq:inequality_2}
\frac{1}{2}\int_{P_l}^{P_r} \mathrm{d} x^1 (\overline{\delta J} ,\overline{\delta J} )-
\frac{1}{2}\int_{A_l}^{A_r} \mathrm{d} x^1 (\overline{\delta J} ,\overline{\delta J} ) \le -\int_{C_l+C_r} \mathrm{d}x^0  \frac{1}{2}(\overline{\delta J} ,\left[\frac{\mathrm{d}x^1}{\mathrm{d}x^0} I -\Lambda\right] \overline{\delta J} ).
\end{equation}
Parametrizing the curves $C_l$ and $C_r$ with $x^0$ and considering the different orientations of the two curves $C_l$ and $C_r$ respect to this parameter
the righthand side of the inequality   \eqref{eq:inequality_2} can be written as
\begin{equation}
\int_0^t \mathrm{d}x^0  \frac{1}{2}(\overline{\delta J} ,\left[\Lambda-\lambda^{(l)} I \right] \overline{\delta J} )
-\int_0^t \mathrm{d}x^0  \frac{1}{2}(\overline{\delta J} ,\left[\Lambda-
\lambda^{(r)} I\right] \overline{\delta J} ).
\end{equation}
The two characteristic curves were chosen such that  $C_l$  has the maximal velocity and $C_r$ the minimal velocity, which corresponds to the extreme eigenvalues of $\Lambda$, 
\begin{equation}
\lambda^{(l)}=\text{max} \frac{( \overline{\delta J} ,\Lambda  \overline{\delta J}) }{( \overline{\delta J} , \overline{\delta J})},\quad\quad\quad \lambda^{(r)}=\text{min} \frac{( \overline{\delta J} ,\Lambda  \overline{\delta J}) }{( \overline{\delta J} , \overline{\delta J})},
\end{equation}
therefore
\begin{equation}
(\overline{\delta J} ,\left[\Lambda-\lambda^{(l)} I \right] \overline{\delta J} )\le 0\quad\quad\quad\text{and}\quad\quad\quad(\overline{\delta J} ,\left[\Lambda-\lambda^{(r)} I \right] \overline{\delta J} )\ge 0.
\end{equation}
Using these inequalities, we obtain
\begin{equation}
\frac{1}{2}\int_{P_l}^{P_r} \mathrm{d} x^1 (\overline{\delta J} ,\overline{\delta J} )\le
\frac{1}{2}\int_{A_l}^{A_r} \mathrm{d} x^1 (\overline{\delta J} ,\overline{\delta J} ). \end{equation}
Restoring the original variables $\delta J =e^{\mu x^0} \overline {\delta J }$ and defining $E(t)$ as
\begin{equation}
\begin{split}
E(t)=\frac{1}{2}\int_{A_l}^{A_r}\mathrm{d}x^1 \delta J^m\delta J^m=
\frac{1}{2}\int_{A_l}^{A_r}\mathrm{d}x^1 w_i^{(m)} w_j^{(m)} \delta \Phi_j \delta \Phi_i , \\
E(0)=\frac{1}{2}\int_{P_1}^{P_k}\mathrm{d}x^1 \delta J^m\delta J^m=
\frac{1}{2}\int_{P_1}^{P_k}\mathrm{d}x^1 w_i^{(m)} w_j^{(m)} \delta \Phi_j \delta \Phi_i , 
\end{split}
\end{equation}
the ``energy inequality'' for this system of linear hyperbolic equations reads
\begin{equation}
0 \le E(t)\le e^{2 \mu t}  E(0).
\end{equation}

\end{document}